\documentclass[fleqn,usenatbib]{mnras}

\usepackage{newtxtext,newtxmath}

\usepackage[T1]{fontenc}

\DeclareRobustCommand{\VAN}[3]{#2}
\let\VANthebibliography\thebibliography
\def\thebibliography{\DeclareRobustCommand{\VAN}[3]{##3}\VANthebibliography}

\usepackage{graphicx}	
\usepackage{amsmath}	

\newcommand{\xhi}{$\langle x_{\rm HI} \rangle$}

\title[Dark pixel fraction from XQR-30]{Updated dark pixel fraction constraints on reionization's end from the 
Lyman-series forests
of XQR-30}

\author[F. B. Davies et al.]{Frederick B. Davies,$^{1}$\thanks{E-mail: davies@mpia.de (FBD)}
Sarah E. I. Bosman,$^{2,1}$
Valentina D'Odorico,$^{3,4,5}$
Sofia Campo,$^{3}$
Andrei Mesinger,$^{4}$\newauthor
Yuxiang Qin,$^{6}$
George D. Becker,$^{7}$
Eduardo Ba{\~n}ados,$^{1}$
Huanqing Chen,$^{8}$
Stefano Cristiani,$^{3,5,9}$    
Xiaohui Fan,$^{10}$\newauthor
Simona Gallerani,$^{4}$
Martin G. Haehnelt,$^{11}$
Laura C. Keating,$^{12}$
Samuel Lai,$^{13}$
Emma Ryan-Weber,$^{14}$
Feige Wang,$^{15}$\newauthor
Jinyi Yang,$^{15}$
and Yongda Zhu$^{10}$
\\
$^{1}$Max-Planck-Institut f\"{u}r Astronomie, K\"{o}nigstuhl 17, D-69117 Heidelberg, Germany\\
$^{2}$Institute for Theoretical Physics, Heidelberg University, Philosophenweg 12, D–69120, Heidelberg, Germany\\
$^{3}$INAF-Osservatorio Astronomico di Trieste, Via Tiepolo 11, I-34143 Trieste, Italy\\
$^{4}$Scuola Normale Superiore, Piazza dei Cavalieri 7, I-56126 Pisa, Italy\\
$^{5}$IFPU-Institute for Fundamental Physics of the Universe, via Beirut 2, I-34151 Trieste, Italy\\
$^{6}$School of Physics, University of Melbourne, Parkville, VIC 3010, Australia\\
$^{7}$Department of Physics \& Astronomy, University of California, Riverside, CA 92521, USA
$^{8}$Augustana Campus, University of Alberta, Camrose, AB T4V2R3, Canada\\
$^{9}$INFN – National Institute for Nuclear Physics, via Valerio 2, I-34127 Trieste, Italy\\
$^{10}$Steward Observatory, University of Arizona, 933 North Cherry Avenue, Tucson, AZ 85721, USA\\
$^{11}$Kavli Institute for Cosmology and Institute of Astronomy, Madingley Road, Cambridge CB3 0HA, UK\\
$^{12}$Institute for Astronomy, University of Edinburgh, Blackford Hill, Edinburgh EH9 3HJ, UK\\
$^{13}$Space \& Astronomy, Commonwealth Scientific and Industrial Research Organisation (CSIRO), P. O. Box 1130, Bentley, WA 6102, Australia\\
$^{14}$Centre for Astrophysics and Supercomputing, Swinburne University of Technology, Hawthorn, Victoria 3122, Australia\\
$^{15}$Department of Astronomy, University of Michigan, 1085 S. University Ave., Ann Arbor, MI 48109, USA\\
}

\date{Accepted XXX. Received YYY; in original form ZZZ}

\pubyear{2025}

\begin{document}
\label{firstpage}
\pagerange{\pageref{firstpage}--\pageref{lastpage}}
\maketitle

\begin{abstract}
The fraction of ``dark pixels'' in the Ly$\alpha$ and other Lyman-series forests at $z\sim5$--$6$ provides a powerful constraint on the end of the reionization process. Any spectral region showing transmission must be highly ionized, while dark regions could be ionized or neutral, thus the dark pixel fraction provides a (nearly) model independent upper limit to the volume-filling fraction of the neutral intergalactic medium, modulo choices in binning scale and dark pixel definition. Here we provide updated measurements of the 3.3\,comoving Mpc dark pixel fraction at $z=4.85$--$6.25$ in the Ly$\alpha$, Ly$\beta$, and Ly$\gamma$ forests of $34$ deep $5.8 \lesssim z\lesssim6.6$ quasar spectra from the (enlarged) XQR-30 sample. Using the negative pixel method to measure the dark pixel fraction, we derive fiducial $1\sigma$ upper limits on the volume-average neutral hydrogen fraction of $\langle x_{\rm HI} \rangle \leq \{0.030+0.048,0.095+0.037,0.191+0.056,0.199+0.087\}$ at $\bar{z}=\{5.481,5.654,5.831,6.043\}$ from the optimally sensitive combination of the Ly$\beta$ and Ly$\gamma$ forests. We further demonstrate an alternative method that treats the forest flux as a mixture of dark and transparent regions, where the latter are modeled using a physically-motivated parametric form for the intrinsic opacity distribution. The resulting model-dependent upper limits on \xhi\ are similar to those derived from our fiducial model-independent analysis. We confirm that the bulk of reionization must be finished at $z>6$, while leaving room for an extended ``soft landing'' to the reionization history down to $z\sim5.4$ suggested by Ly$\alpha$ forest opacity fluctuations. 
\end{abstract}




\section{Introduction}\label{sec:intro}

The reionization of hydrogen in the intergalactic medium (IGM) was a pivotal moment in the history of the Universe, wherein the first stars and galaxies exerted their influence on cosmological scales, transforming the intergalactic gas into the ionized plasma that persists today. Observational probes of the reionization epoch are largely limited to three general categories: scattering of the cosmic microwave background (CMB) off of free electrons, emission and absorption by the hyperfine 21\,cm transition of neutral hydrogen, and scattering by Ly$\alpha$ and other resonant Lyman-series transitions of neutral hydrogen.

The CMB provides an integral constraint on the reionization process via the electron-scattering optical depth $\tau_{\rm e}$ (e.g.~\citealt{Zaldarriaga97}), representing the integral of ionized gas from the present-day to the surface of last scattering at $z\sim1100$. The latest measurements by the \emph{Planck} mission suggest a characteristic reionization epoch of $z=7.67\pm0.73$ \citep{Planck18}, with some additional systematic uncertainty depending on how the reionization history is allowed to vary\footnote{Or which CMB data are used in the measurement. Recent work suggests a larger $\tau_e$ and thus earlier reionization could still be viable if low-$l$ CMB data are excluded (\citealt{Sailer25,Jhaveri25}; although see \citealt{Cain25tau}).}. Further information on the duration of the process can be determined from the patchy kinetic Sunyaev-Zel'dovich (kSZ) effect (e.g.~\citealt{Aghanim96,Gruzinov98,Knox98}), which primarily depends on the duration of the reionization process (e.g.~\citealt{Zahn05,McQuinn05,Mesinger12,Battaglia13}), but so far the patchy kSZ signal is only detected by CMB experiments at a signal-to-noise ratio of $\sim3$ \citep{Reichardt21}. The 21\,cm signal is another promising probe of the reionization process; the large-scale fluctuations in the reionization topology are, in principle, detectable in emission or absorption directly from the neutral hydrogen in the IGM itself \citep{Madau97,Furlanetto06}. Due to substantial foregrounds and instrumental challenges, a detection of the signal remains elusive, but the upper limits achieved thus far (e.g.~\citealt{HERA23,Mertens25,Ceccotti25,Nunhokee25}) nevertheless provide valuable astrophysical constraints (e.g.~\citealt{HERA23,Bevins24,Sims25,Ghara25,Nunhokee25}).

The Ly$\alpha$ forest in the spectra of high-redshift quasars provides a sensitive view of the neutral hydrogen content of the early Universe. Before reionization has completed, the neutral IGM will imprint completely saturated Ly$\alpha$ absorption known as the Gunn-Peterson trough \citep{GP65}. Indeed, early observations of the first quasars discovered above redshift 6 showed dark troughs in both the Ly$\alpha$ and Ly$\beta$ forests \citep{Becker01,White03}. While the absence of \emph{ubiquitous} troughs at $z<6$ was initially interpreted to mean that reionization was complete (e.g. \citealt{Fan06}, although see \citealt{Mesinger10}), it was eventually realized that the large observed scatter in Ly$\alpha$ forest opacity \citep{Becker15} was difficult to explain in the context of complete reionization (although see \citealt{D'Aloisio15,DF16,Chardin17}). In contrast, models of patchy, incomplete reionization at $z<6$ provided an easier and better-fitting explanation \citep{Kulkarni19,ND20,Keating19,Choudhury21,Qin21}. The most recent measurements of large-scale Ly$\alpha$ forest fluctuations suggest that these excess fluctuations persist all the way down to $z\sim5.3$ \citep{Bosman22}. 

However, in some respects Ly$\alpha$ absorption is \emph{too} sensitive -- H\,{\small I} fractions of $\sim10^{-4}$ are sufficient for nearly complete saturation \citep{GP65}. Fortunately, the large H\,{\small I} column densities of a truly neutral IGM are sufficient for their Lorentzian damping wings to imprint substantial absorption at large velocity separations \citep{ME98}. Provided that the presence of non-IGM damped Ly$\alpha$ absorbers in the foreground can be ruled out \citep{Simcoe12,Banados19,Davies25PDLA}, this long-range damping wing absorption enables additional \xhi\ constraining power from sensitive rest-frame Ly$\alpha$ spectroscopy of high-redshift quasars (e.g.~\citealt{MH04,Greig17b,Greig24,Davies18b,Durovcikova24,Kist25}), gamma-ray bursts (e.g.~\citealt{Totani06,Chornock13,Lidz21,Fausey25}), and galaxies (e.g.~\citealt{Umeda25,Mason25}), as well as the statistics of Ly$\alpha$ emission lines in high-redshift galaxies (e.g.~\citealt{Mesinger15,Mason18,Bolan22,Nakane24,Kageura25}). Hints of damping wing absorption from neutral islands deeper within the Ly$\alpha$ forest \citep{ML15} have also begun to emerge \citep{Spina24,Zhu24,Becker24} although its interpretation may be non-trivial \citep{Gnedin25DW}.

The Ly$\alpha$ absorption constraints mentioned above are inherently \emph{model-dependent}, that is, they rely on calibration with mock data extracted from cosmological simulations to connect observations to physical properties of the IGM, i.e. the neutral fraction. However, given the broad uncertainties that remain in our knowledge of the reionization process, and the expensive nature of detailed reionization simulations, the derived \xhi\ constraints depend on the specific model employed and are difficult to compare to each other. The work of \citet{Mesinger10} emphasized that more reliable constraints could be derived from \emph{model-independent} probes, and devised one such method using the Ly$\alpha$ and Ly$\beta$ forests. They considered the fact that any spectral region of the Ly$\alpha$ or Ly$\beta$ forest with detected transmission \emph{must be} highly ionized, while any region without transmission \emph{could be} neutral. Thus the fraction of path length with zero transmission provides an upper limit to \xhi. The method was first tested empirically in \citet{McGreer11}, where to maximize sensitivity they adopted a spectral binning scale of 3.3 comoving Mpc, although this choice introduces a small degree of model dependence. The resulting constraints on \xhi\ by \citet{McGreer11} were relatively weak, but deeper spectroscopic measurements enabled a subsequent work by \citet{McGreer15} to place tight upper limits of $\langle x_{\rm HI} \rangle \leq 0.04+0.05\ (0.06+0.05)$ ($1\sigma$) at $z=5.6\ (5.9)$. More recently, \citet{Jin23} employed a similar technique on a sample of much higher redshift quasars to place the first \xhi\ upper limits at $z>6.2$: $\langle x_{\rm HI} \rangle \leq 0.79+0.04$, $\leq0.87+0.03$, and $\leq0.94+0.06$ at $z=6.3$, $6.5$, and $6.7$, respectively.

In this work, we measure the dark pixel fraction using deep VLT/X-Shooter spectroscopy from the XQR-30 survey \citep{D'Odorico23}, plus additional deep spectroscopy of similar quality in public archives (E-XQR-30). We employ methods similar to \citet{McGreer11,McGreer15} to derive upper limits on \xhi\ using the Ly$\alpha$, Ly$\beta$, and (for the first time) Ly$\gamma$ forests. Finally, we also investigate a more model-dependent parametric approach, treating the forest as a mixture of ``transparent'' and ``dark'' pixels, and find very consistent results to our fiducial model-independent analysis. 

We assume a flat $\Lambda$CDM cosmology consistent with \citet{Planck18}, adopting the cosmological parameter values $h=0.68$, $\Omega_m=0.31$, and $\Omega_\Lambda=0.69$. Distance units are comoving unless specified otherwise.

\section{XQR-30 Lyman-series Data Set}\label{sec:data}

\begin{figure}
\begin{center}
\resizebox{8.5cm}{!}{\includegraphics[trim={1em 1em 1em 1em},clip]{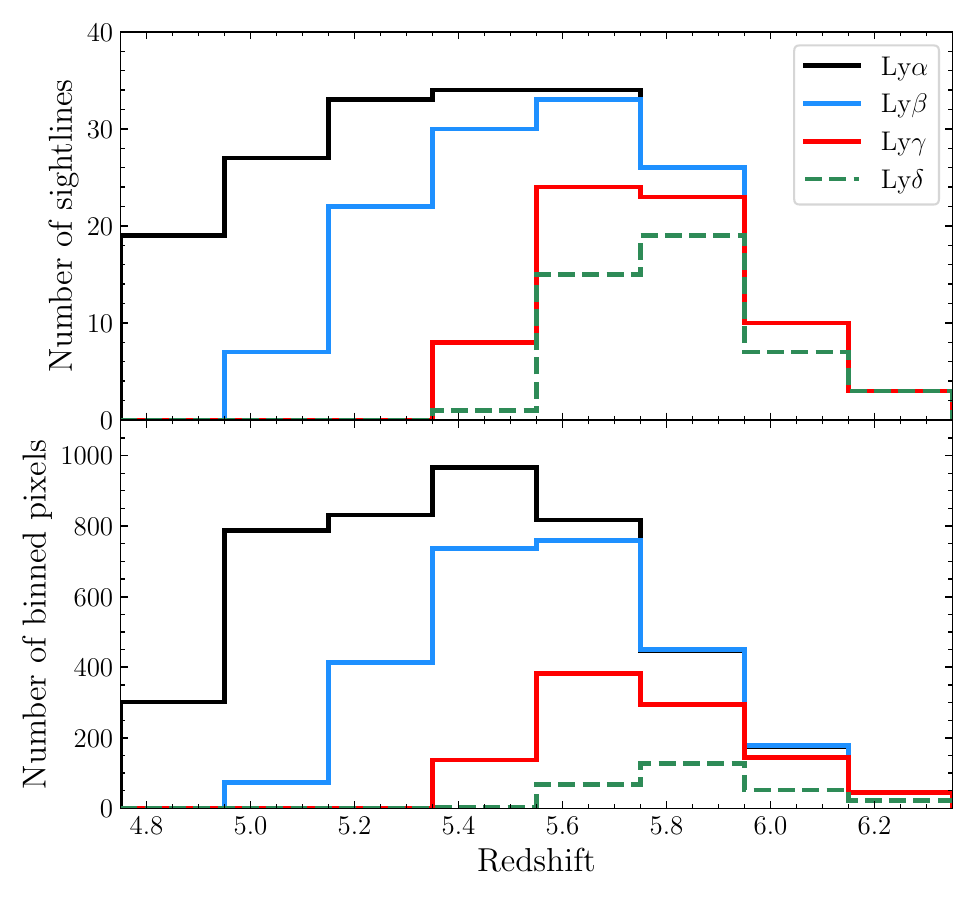}}\\
\end{center}
\caption{Top: Number of quasar sightlines covering the Ly$\alpha$ (black), Ly$\beta$ (blue), and Ly$\gamma$ (red) forests within the eight $\Delta z=0.2$ redshift bins we consider in this work. Bottom: Corresponding number of binned pixels in each forest/redshift bin combination. The green dashed lines in both panels show the corresponding numbers for the Ly$\delta$ forest, discussed in Appendix~\ref{app:delta}.}
\label{fig:npix}
 \end{figure}

We use data drawn from the XQR-30 large program (1103.A-0817(A); \citealt{D'Odorico23}) combined with literature data of comparable quality -- the so-called ``enlarged'' or E-XQR-30 data set \citep{D'Odorico23}. These data consist of deep VLT/X-Shooter \citep{Vernet11} spectroscopy of 30+12 quasars with redshifts $5.8 \lesssim z \lesssim 6.6$, which have been reduced with a custom data reduction pipeline (see \citealt{D'Odorico23} for details). After careful spectral quality cuts (described below and in Appendix~\ref{app:parity}), and removing the most prominent broad absorption-line (BAL) quasars, in our analysis we use 34 out of 42 of the quasars (Table~\ref{tab:qsos}). We show a summary of the number of sightlines and binned Lyman-series pixels (see below) in Figure~\ref{fig:npix}.

\begin{table*}\centering
\begin{tabular}{l c c c l}
Name & $z_{\rm q}$ & $z_{\rm q}$ ref. & Median $\tau_{\rm eff,lim}$ ($1\sigma$) & Notes \\
\hline \hline
PSO J231$-$20     & 6.5869 & 4 & 5.533 & XQR-30; BAL \\
PSO J183$+$05     & 6.4386 & 4 & 5.594 & XQR-30; Abs @ $z=6.4041,6.0642$ \\
DELS J1535$+$1943 & 6.4378 & 9 & 4.942 & XQR-30 \\
VDES J2211$-$3206 & 6.342  & 9 & 5.194 & XQR-30; BAL \\
\hline
ATLAS J025$-$33   & 6.3373 & 4 & 6.173 & E-XQR-30\\
SDSS J0100$+$2802 & 6.3269 & 4 & 7.615 & E-XQR-30; Abs @ $z=6.1434,6.1114,5.9450,5.7974$ \\
SDSS J1030$+$0524 & 6.309  & 8 & 5.546 & E-XQR-30 \\
PSO J217$-$07     & 6.1993 & 8 & 5.256 & XQR-30; BAL \\
PSO J065$-$26     & 6.1871 & 4 & 6.042 & XQR-30; Abs @ $z=6.1235,5.8677$ \\
\hline
PSO J060$+$24     & 6.1793 & 9 & 5.414 & XQR-30; Abs @ $z=5.6993$ \\
PSO J217$-$16     & 6.1466 & 9 & 5.736 & XQR-30 \\
ULAS J1319$+$0950 & 6.1347 & 4 & 5.889 & E-XQR-30; Abs @ $z=6.0172$ \\
PSO J239$-$07     & 6.1102 & 5 & 5.700 & XQR-30; Abs @ $z=5.9918$; BAL \\
SDSS J0842$+$1218 & 6.0754 & 4 & 5.776 & XQR-30 \\
\hline
PSO J158$-$14     & 6.0685 & 5 & 5.569 & XQR-30; Abs @ $z=5.8986$ \\
SDSS J1306$+$0356 & 6.0330 & 4 & 5.499 & E-XQR-30 \\
VDES J0408$-$5632 & 6.0264 & 9 & 5.786 & XQR-30 \\
ATLAS J029$-$36   & 6.021  & 8 & 5.401 & XQR-30 \\
PSO J009$-$10     & 6.0040 & 4 & 4.803 & XQR-30; BAL \\
\hline
SDSS J2310$+$1855 & 6.0031 & 2 & 6.022 & XQR-30; Abs @ $z=5.9388$ \\
PSO J007$+$04     & 6.0015 & 4 & 5.413 & XQR-30; Abs @ $z=5.9917$ \\
SDSS J0818$+$1722 & 5.9991 & 9 & 6.191 & E-XQR-30; Abs @ $z=5.8767,5.7912$ \\
VDES J2250$-$5015 & 5.9988 & 9 & 5.050 & XQR-30; BAL \\
ULAS J0148$+$0600 & 5.9896 & 9 & 6.290 & E-XQR-30 \\
\hline
PSO J029$-$29     & 5.984  & 8 & 5.519 & XQR-30 \\
PSO J108$+$08     & 5.9647 & 9 & 5.859 & XQR-30; Abs @ $z=5.5624$ \\
PSO J183$-$12     & 5.917  & 8 & 5.803 & XQR-30 \\
PSO J242$-$12     & 5.8468 & 9 & 5.259 & XQR-30 \\
PSO J023$-$02     & 5.8464 & 9 & 4.995 & XQR-30; Abs @ $z=5.4869$; BAL \\
\hline
PSO J025$-$11     & 5.8414 & 9 & 5.412 & XQR-30; Abs @ $z=5.8385,5.7763$ \\
PSO J065$+$01     & 5.8348 & 9 & 5.188 & XQR-30 \\
SDSS J0836$+$0054 & 5.804  & 8 & 5.789 & E-XQR-30 \\
PSO J308$-$27     & 5.803  & 9 & 5.463 & XQR-30; Abs @ $z=5.6268,5.4400$ \\
SDSS J0927$+$2001 & 5.7722 & 1 & 5.605 & E-XQR-30 \\
\hline \hline
Excluded from analysis & & & \\
\hline
DELS J0923$+$0402 & 6.6330 & 7 & -- & XQR-30; Extreme BAL \\
PSO J323$+$12     & 6.5872 & 4 & -- & XQR-30; Positive zero-level bias\\
PSO J036$+$03     & 6.5405 & 4 & -- & E-XQR-30; Negative zero-level bias \\
VDES J0224$-$4711 & 6.5223 & 9 & -- & XQR-30; Positive zero-level bias \\
UHS J0439$+$1634  & 6.5188 & 6 & -- & E-XQR-30; Contamination by lensing galaxy \\
PSO J359$-$06     & 6.1722 & 5 & -- & XQR-30; Positive zero-level bias \\
CFHQS J1509$-$1749 & 6.1228 & 3 & -- & E-XQR-30; Positive zero-level bias \\
PSO J089$-$15     & 5.9756 & 9 & -- & XQR-30; Extreme BAL \\
\hline
\end{tabular}
\caption{List of quasar spectra considered in this work. The columns from left to right show the quasar name, quasar systemic redshift $z_{\rm q}$, the corresponding reference to the systemic redshift, the median limiting effective optical depth $\tau_{\rm eff,lim}$ (see text), and additional quasar-specific notes. The median $\tau_{\rm eff}$ limits are computed over the whole spectral range considered in this work. Absorbers noted by ``Abs'' are O\,{\small I} absorbers from \citet{Sodini24}. Systemic redshift references: 1 -- \citet{Carilli07}, 2 -- \citet{Wang13}, 3 -- \citet{Decarli18}, 4 -- \citet{Venemans20}, 5 -- \citet{Eilers21E}, 6 -- \citet{Yang19}, 7 -- \citet{Yang21}, 8 -- \citet{Bosman22}, 9 -- Neelemann et al., in prep. }
\label{tab:qsos}
\end{table*}

\begin{figure*}
\begin{center}
\resizebox{16cm}{!}{\includegraphics[trim={1em 1em 1em 1em},clip]{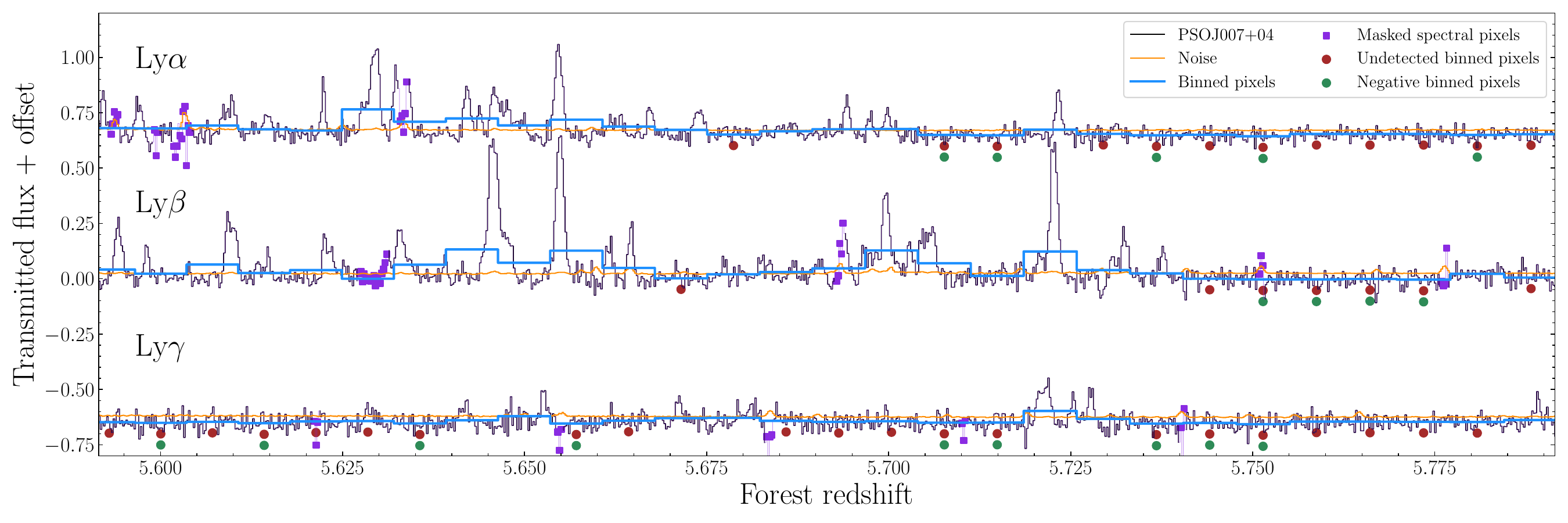}}\\
\end{center}
\caption{Example of a masked and binned quasar spectrum from our data set (PSO~J007$+$04, $z_{\rm q}=6.0015$). Brown and green points below the spectrum indicated undetected (${\rm S/N}<2$) and negative binned pixels, respectively. The forests Ly$\alpha$/Ly$\gamma$ forest transmissions are shown offset by $\pm0.65$ for clarity.}
\label{fig:spec}
 \end{figure*}

Despite the high signal-to-noise of the spectra, some artifacts remain, particularly in the vicinity of sky emission lines. We clean the spectra in two steps. First, we mask pixels which are overly negative with ${\rm S/N}<-3$, corresponding to $\simeq0.2\%$ of pixels. Second, we mask pixels whose noise values are above the 97.5-percentile of the distribution within the relevant wavelength range. This latter step is performed to remove sky emission line features uniformly from each spectrum in lieu of individually tuned peak-finding approaches (e.g.~\citealt{Zhu22}), and we found that our results are insensitive to the exact percentile chosen. We note that for the brightest quasars in our sample, the pixel noise can include a non-negligible contribution from the object flux, potentially leading to masking of the brightest transmission spikes at $z\lesssim5.5$ and biasing the dark fraction high, however in practice we find that only sky-dominated pixels are masked.
Finally, we expand these masked regions by 2 spectral pixels ($\sim20$\,km/s) in either direction to clean up any residual noise correlations. We show an example of the outcome of this procedure in Figure~\ref{fig:spec}, where the lightly-colored (unbinned) pixels are masked.

We also inspect each spectrum for observational artifacts that may contaminate the zero-level, with potentially catastrophic consequences for measurements close to the noise level. Specifically, we investigate the positive-negative ``parity'' of each spectrum with two tests: the cumulative parity, and the binned parity. We first transform each spectrum into $\pm1$ values, $F_\pm$, corresponding to the sign of the flux in each pixel. In the cumulative parity test, we inspect $\sum{F_\pm}(<\lambda)$ as a function of observed wavelength $\lambda$ within the X-Shooter/VIS portion of the quasar spectrum ($\lambda>5500$\,\AA). At wavelengths blueward of the Lyman limit ($\lambda_{\rm rest}<911.76$\,\AA) the spectrum should be roughly zero due to the short mean free path of ionizing photons at $z>5.8$ \citep{Becker21,Zhu23}, thus we expect the cumulative parity to be roughly flat at blue wavelengths, followed by discrete jumps due to flux spikes encountered at redder wavelengths. An incorrect zero-level in the spectrum will appear as an overall positive or negative drift across the entire spectrum irrespective of the location of real features in the flux. We also inspect the binned parity across the spectrum in $\sim300$\,km/s bins and look for statistically significant ($>2\sigma$) local (negative) deviations in parity that may be indicative of poor sky subtraction. Inspection of these two metrics for every quasar resulted in the exclusion of six objects, five with excess positive parity (PSO~J323$+$12, VDES~J0224$-$4711, UHS~J0439$+$1634\footnote{The positive drift in this strongly-lensed quasar is due to blending with a foreground lensing galaxy \citep{Fan19}.}, PSO~J359$-$06, CFHQS~J1509$-$1749) and one with excess negative parity (PSO~J036$+$03). Examples of the parity tests for excluded and non-excluded quasars can be found in Appendix~\ref{app:parity}.

We aim to study the dark pixel fraction of the Ly$\alpha$ ($\lambda_\alpha=1215.67$\,\AA), Ly$\beta$ ($\lambda_\beta=1025.72$\,\AA), and (for the first time) Ly$\gamma$ ($\lambda_\gamma=972.54$\,\AA) forests of the E-XQR-30 quasars. Traditionally, the Ly$\alpha$ forest extends from Ly$\alpha$ at the redshift of the background quasar $z_{\rm q}$ to the start of the Ly$\beta$ forest. We exclude the region in the vicinity of the quasar, i.e. the proximity zone, by starting the Ly$\alpha$ forest at $z_{\rm q}-\Delta z_{\rm prox}$, where we choose $\Delta z_{\rm prox}=0.2$. This is a relatively large exclusion compared to the $\Delta z_{\rm prox}=0.1$ of \citet{McGreer11,McGreer15}, but for most quasars it is slightly smaller than the exclusion by \citet{Jin23}, who started the Ly$\alpha$ forest at $\lambda = 1176$\,\AA $\times (1+z_{\rm q})$. While the Ly$\alpha$ regions with by-eye elevated transmission are often much smaller than our exclusion, we conservatively opt for this larger exclusion zone as it encompasses the largest proximity zone in our sample (towards SDSS~J0100+2802), and there may be second-order effects in the ionizing background to larger distances \citep{Davies20ghost}. The short wavelength end of the Ly$\alpha$ forest of each quasar is then cut off at the onset of Ly$\beta$ forest absorption ($\lambda = \lambda_\beta \times (1+z_{\rm q})$), such that the majority of the quasars studied probe Ly$\alpha$ redshifts down to $z_{\rm Ly\alpha}\sim5$. For the Ly$\beta$ and Ly$\gamma$ forests, we apply the same proximity zone exclusion, but we treat the low redshift end differently, probing all the way down to $\lambda_{\rm rest}=915$\,\AA\footnote{We expect wavelengths at $\lambda_{\rm rest}<912$\,\AA\ to experience additional Lyman limit absorption, and leave a buffer in case the quasar redshifts are not precise enough (e.g. those derived from the Mg\,{\small II} emission line). Our results are insensitive to the exact limit chosen.}. The addition of such heavily foreground-contaminated pixels may be counterintuitive -- the individual Ly$\beta$ and Ly$\gamma$ forests will have a much higher observed fraction of dark pixels in regions contaminated by foreground absorption. But note that when the Lyman-series forests are considered in \emph{combination}, any pixels with transmitted flux not present in lower-order lines will rule out a truly dark pixel (corresponding to neutral IGM) at that location. We find that this hypothesis holds in practice, with greatly improved constraining power from the larger path length allowed by the extended wavelength coverage.

Dense absorbers with large H\,{\small I} column density can also introduce large correlated regions with zero transmission, which may bias the dark fractions to higher values. We mask regions corresponding to potential high column density Damped Ly$\alpha$ (DLA) systems at the locations of the \citet{Sodini24} O\,{\small I} metal absorbers\footnote{\citet{Sodini24} searched the E-XQR-30 spectra for O\,{\small I} absorbers by eye. No additional O\,{\small I} absorbers were found in the complete automated metal absorber catalogue from \citet{RDavies23}.} with an exclusion width of $\Delta v=1500$\,km/s in the Ly$\alpha$ forest. For the other Lyman-series forests, we conservatively scale this $\Delta v$ by $\ln{f_{\rm line}\lambda_{\rm line}}$, where $f_{\rm line}$ is the oscillator strength of the transition at rest-frame wavelength $\lambda_{\rm line}$, corresponding to the flat part of the curve of growth. The Lyman-series forests for each metal absorber were manually inspected, with some showing significant transmission within the nominal mask region. For these absorbers, we reduced the size of the mask for the corresponding transition and all higher order transitions to the implied maximum width. In the end, roughly $\approx2.4\%$ of spectral pixels are masked, with a similar few percent decrease in our final dark fractions.

Similar to \citet{Bosman22}, we mask regions corresponding to strong BAL absorption as determined by \citet{Bischetti22}. Specifically, we mask the Si\,{\small IV}, N\,{\small V}, O\,{\small VI}, and Lyman-series lines at velocities with $>20\%$ absorption in the C\,{\small IV} line. While these regions may still show transmission in the Lyman-series forests as long as the BAL troughs are not fully saturated, their effective signal-to-noise will be much lower and difficult to quantify, so we conservatively mask them from the analysis. We find that this masking has only a minor effect on our final results, reducing the final dark fraction measurements by $\sim1-2\%$. Two quasars in the E-XQR-30 data set, DELS~J0923$+$0402 and PSO~J089$-$15, exhibit such strong and broad BAL absorption that their spectra are almost fully contaminated, thus we exclude them from our analysis.

\begin{figure}
\begin{center}
\resizebox{8.5cm}{!}{\includegraphics[trim={1em 1em 1em 1em},clip]{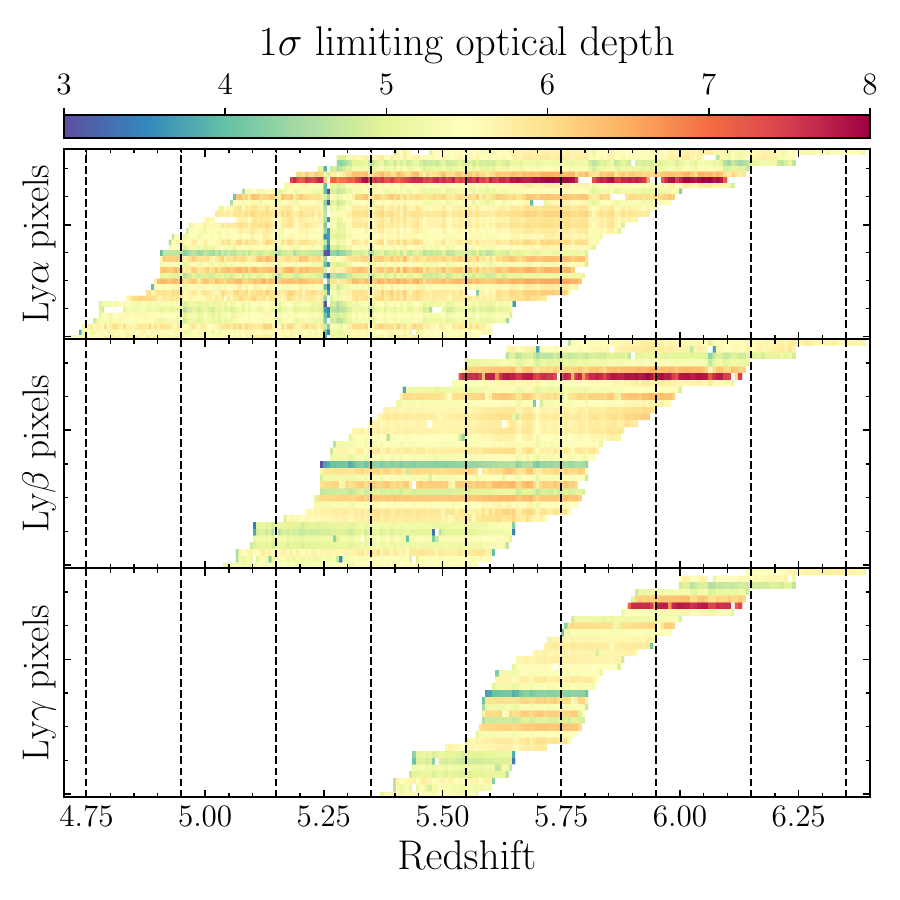}}\\
\end{center}
\caption{Limiting effective optical depths $\tau_{\rm eff,lim}=-\ln{\sigma_{\rm bin}/F_{\rm cont}}$ along each quasar sightline for binned pixels in the Ly$\alpha$ (top), Ly$\beta$ (middle), and Ly$\gamma$ (bottom) forests. The vertical stripe of low $\tau_{\rm eff,lim}$ at $z\simeq5.25$ in the Ly$\alpha$ forest is due to strong telluric absorption from the O$_2$ A-band at $\sim7600$\,\AA. Note also the spectrum of SDSS\,J0100$+$2802 which has extraordinarily high signal-to-noise, resulting in the horizontal stripes of particularly high $\tau_{\rm eff,lim}$.}
\label{fig:taulim}
 \end{figure}

To estimate the sensitivity of our spectra to IGM transmission, we compute the limiting effective optical depth $\tau_{\rm eff,lim}$ from \citet{McGreer15}, corresponding to $-\ln{\sigma_{\rm bin}/F_{\rm cont}}$, where $\sigma_{\rm bin}$ is the flux uncertainty after binning to a scale of 3.3\,Mpc (see Section~\ref{sec:bin} below), and $F_{\rm cont}$ is the unabsorbed quasar continuum. To estimate the quasar continuum, we adopt a simple broken power-law model with $F_\lambda \propto \lambda^{-1.5}$ normalized to $\lambda_{\rm rest}=1290$\,\AA, inflecting to $F_\lambda \propto \lambda^{-0.5}$ at $\lambda_{\rm rest}<1000$\,\AA. More accurate continuum predictions are available at Ly$\alpha$ forest wavelengths for most of the sample \citep{Bosman21,Bosman22}, but we opt for a more simplistic model that can be applied more uniformly and to shorter wavelengths covering the higher order Lyman series forests. The exact continuum level is not important for our fiducial model-independent analyses in Section~\ref{sec:indep}, and as will be discussed below in Section~\ref{sec:dep}, likely only has a small impact on our proof-of-concept model-dependent analysis. In Figure~\ref{fig:taulim} we show the resulting $\tau_{\rm eff,lim}$ as a function of wavelength in the Ly$\alpha$, Ly$\beta$, and Ly$\gamma$ forests, and report the median value for quasar spectrum across the whole considered IGM wavelength range in Table~\ref{tab:qsos}.

\section{Model-independent IGM neutral fraction constraints}\label{sec:indep}

We derive constraints on the IGM neutral fraction in a similar fashion to \citet[][see also \citealt{Jin23}]{McGreer11,McGreer15} by using the so-called ``dark pixel fraction'' technique, originally proposed by \citet{Mesinger10}. The idea is simple: any segment of the Ly$\alpha$ (or other Lyman series) forest with detected transmission must be highly ionized \citep{GP65}, while segments that are dark \emph{may be} neutral. The fraction of the forest covered by dark segments (or one minus the fraction covered by transparent segments) then provides an upper limit to the IGM neutral fraction. In detail, two choices must be made -- first, how are the segments defined, and second, how to identify which segments are dark.

\begin{figure*}
\begin{center}
\resizebox{8.0cm}{!}{\includegraphics[trim={1.5em 1em 1em 1em},clip]{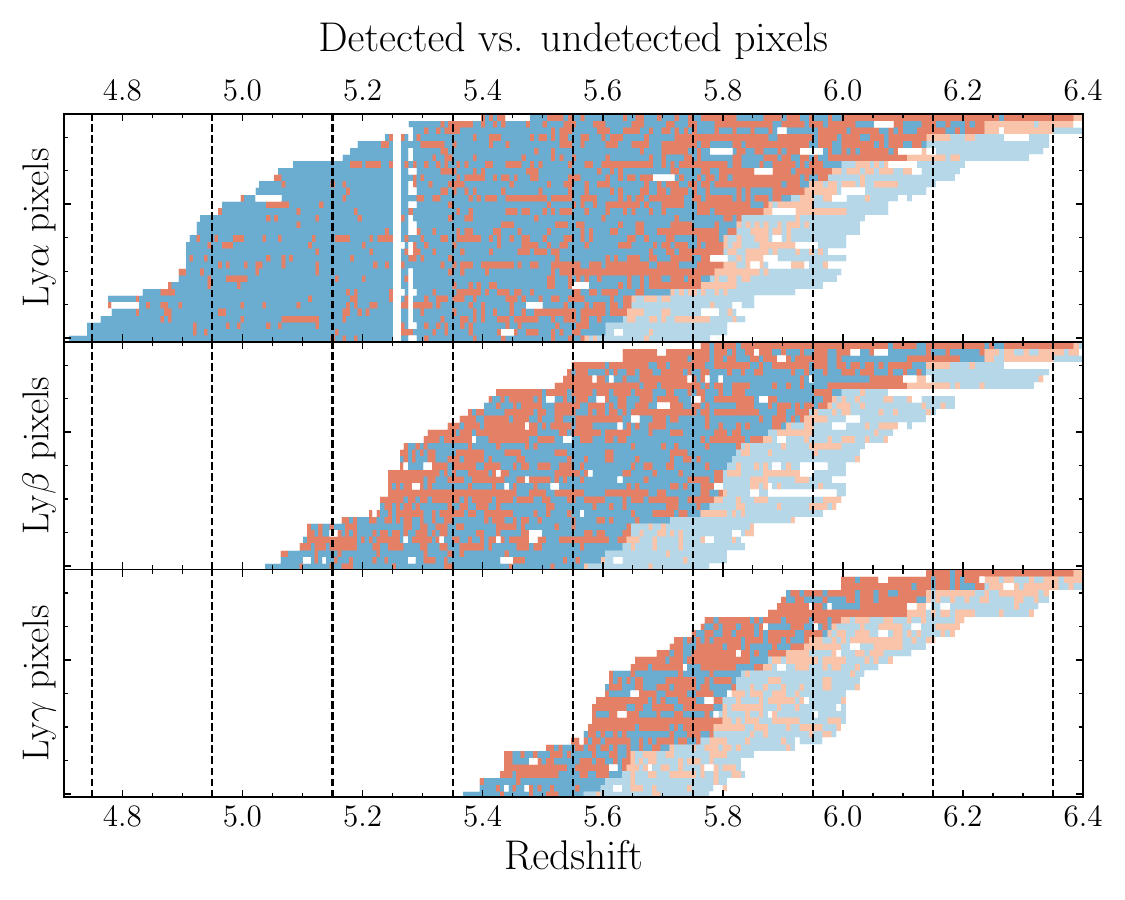}}
\resizebox{8.0cm}{!}{\includegraphics[trim={1.0em 1em 1em 1em},clip]{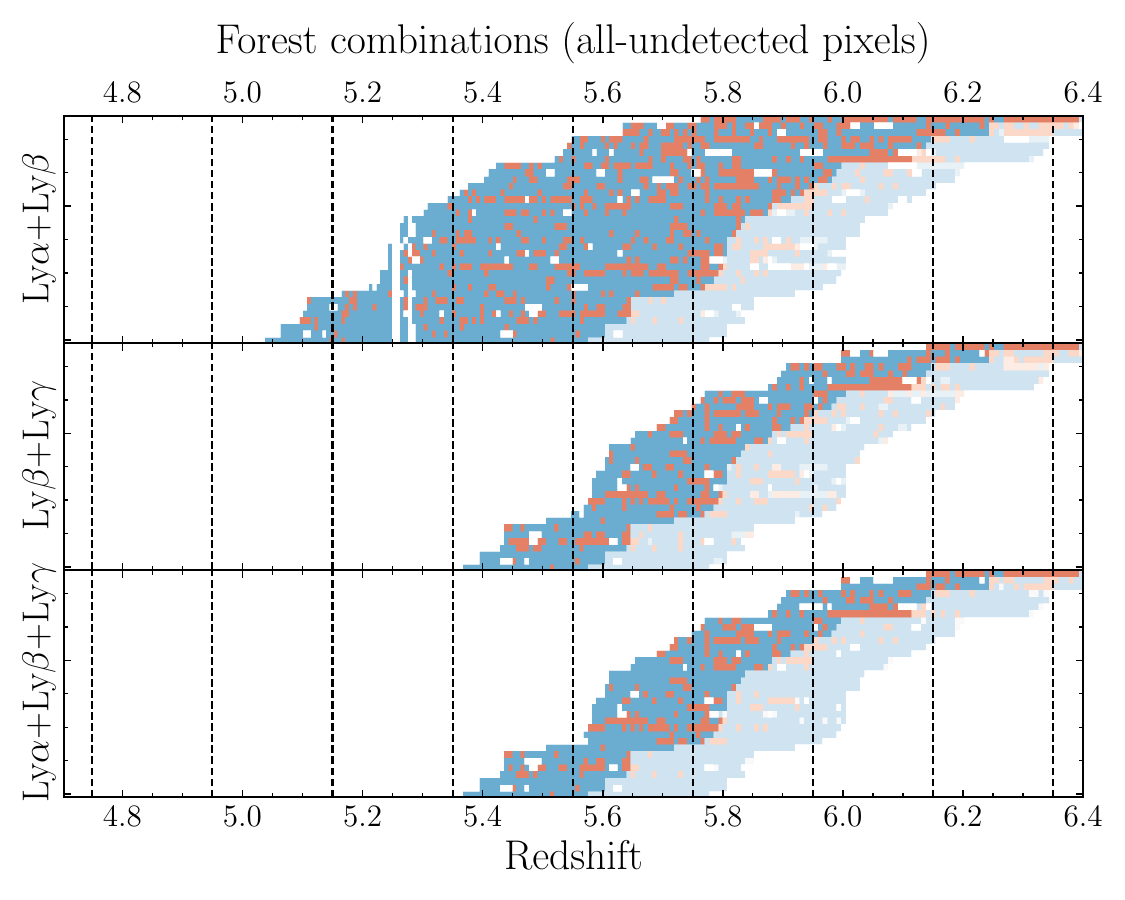}}\\
\end{center}
\vspace{-1.0em}
\caption{Distribution of detected (${\rm S/N}>2$, blue) and undetected (${\rm S/N}<2$, red) binned pixels in the $35$ E-XQR-30 quasar sightlines used in this work. White binned pixels within the low and high redshift boundaries of each quasar sightline have more than 50\% of their spectral pixels masked. The regions with faded color at the high redshift end of each sightline show pixels within the quasar proximity zone exclusion. Left: Binned pixels in the Ly$\alpha$ (top), Ly$\beta$ (middle), and Ly$\gamma$ forests. Right: Forest combinations Ly$\alpha$$+$Ly$\beta$ (top), Ly$\beta$$+$Ly$\gamma$ (middle), and Ly$\alpha$$+$Ly$\beta$$+$Ly$\gamma$ (bottom), where blue pixels are detected in any or all forests while red pixels are undetected in all forests.}
\label{fig:threshpix}
 \end{figure*}

\begin{figure*}
\begin{center}
\resizebox{11cm}{!}{\includegraphics[trim={1.0em 1em 1em 0.5em},clip]{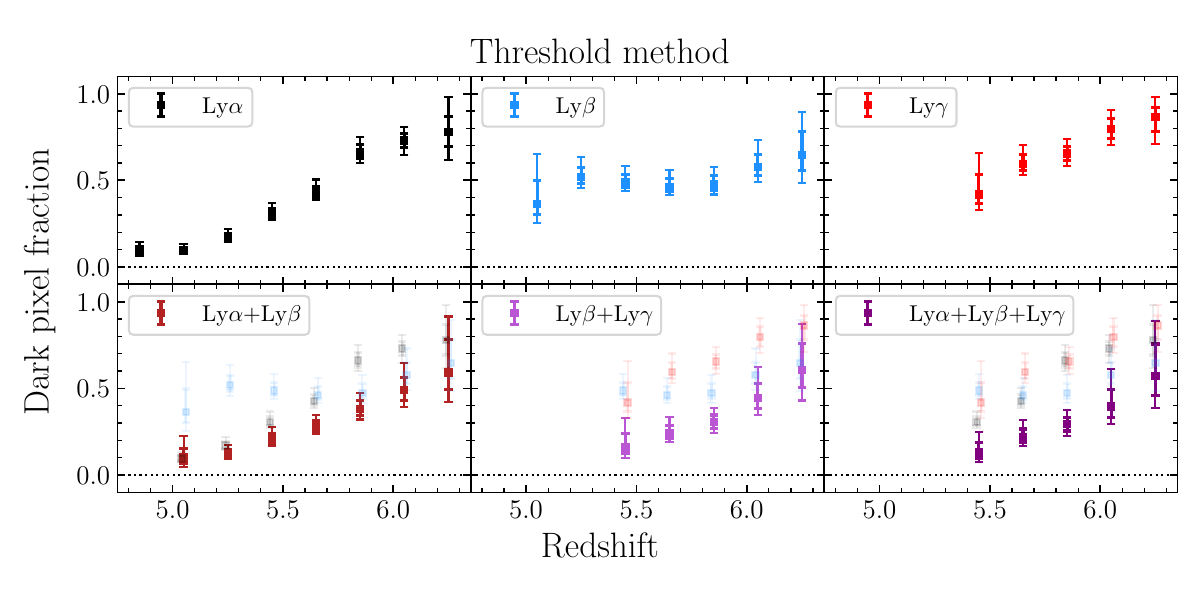}}
\end{center}
\vspace{-2.0em}
\caption{Measurements of the dark pixel fraction in each Lyman-series forest and forest combination inferred from the threshold method applied to the E-XQR-30 data set. Uncertainties (thick: 1$\sigma$, thin: 2$\sigma$) are derived from bootstrap resampling of the quasar sightlines contributing to each bin, or from the confidence interval of the binomial fraction, whichever is larger. For the forest combinations (lower panels), we show the dark pixel fractions of the individual contributing forests in a light color with slight redshift offsets for clarity.}
\label{fig:dark_thresh}
 \end{figure*}

\begin{table*}\centering
\begin{tabular}{c c c c c c c c c}
Line & $z$ range & $\bar{z}$ & $N_{\rm LOS}$ & $N_{\rm pix}$ & $N_{\rm un-det}$ & $f_{\rm dark}^{\rm thresh}+1\sigma(+2\sigma)$ & $N_{\rm neg}$ & $f_{\rm dark}^{\rm neg}+1\sigma(+2\sigma)$ \\
\hline \hline
Ly$\alpha$ & 4.75--4.95 & 4.875 & 19 & 302 & 28 & $\mathbf{0.095+0.023(+0.052)}$ & 9 & $\mathbf{0.060+0.023(+0.053)}$ \\
& 4.95--5.15 & 5.056 & 27 & 788 & 73 & $\mathbf{0.095+0.018(+0.038)}$ & 25 & $\mathbf{0.063+0.017(+0.036)}$ \\
& 5.15--5.35 & 5.252 & 33 & 832 & 139 & $0.171+0.023(+0.048)$ & 49  & $0.118+0.022(+0.048)$ \\
& 5.35--5.55 & 5.448 & 34 & 966 & 289 & $0.306+0.031(+0.064)$ & 115 & $0.238+0.026(+0.053)$ \\
& 5.55--5.75 & 5.641 & 34 & 818 & 343 & $0.429+0.037(+0.079)$ & 166 & $0.406+0.051(+0.103)$ \\
& 5.75--5.95 & 5.831 & 26 & 447 & 289 & $0.661+0.045(+0.088)$ & 127 & $0.568+0.071(+0.145)$ \\
& 5.95--6.15 & 6.037 & 10 & 175 & 125 & $0.731+0.037(+0.074)$ & 53  & $0.606+0.072(+0.146)$ \\
& 6.15--6.35 & 6.225 & 3  & 46  & 35  & $0.778+0.089(+0.202)$ & 15  & $0.652+0.144(+0.294)$ \\
\hline Ly$\beta$ & 4.95--5.15 & 5.112 & 7 & 73 & 26 & $0.364+0.131(+0.275)$ & 10 & $0.274+0.097(+0.201)$ \\
& 5.15--5.35 & 5.271 & 22 & 413 & 209 & $0.518+0.057(+0.116)$ & 73  & $0.354+0.064(+0.136)$ \\
& 5.35--5.55 & 5.450 & 30 & 736 & 351 & $0.488+0.050(+0.095)$ & 128 & $0.348+0.045(+0.089)$ \\
& 5.55--5.75 & 5.642 & 33 & 760 & 344 & $0.463+0.048(+0.097)$ & 119 & $0.313+0.051(+0.104)$ \\
& 5.75--5.95 & 5.832 & 26 & 450 & 209 & $0.475+0.054(+0.104)$ & 85  & $0.378+0.065(+0.130)$ \\
& 5.95--6.15 & 6.038 & 10 & 179 & 101 & $0.577+0.078(+0.161)$ & 37  & $0.413+0.064(+0.133)$ \\
& 6.15--6.35 & 6.225 & 3  & 46  & 29  & $0.645+0.135(+0.250)$ & 14 & $\mathbf{0.609+0.143(+0.293)}$ \\
\hline Ly$\gamma$ & 5.35--5.55 & 5.480 & 8 & 137 & 56 & $0.418+0.115(+0.242)$ & 18 & $0.263+0.073(+0.169)$ \\
& 5.55--5.75 & 5.654 & 24 & 382 & 222 & $0.594+0.053(+0.106)$ & 80 & $0.419+0.064(+0.127)$ \\
& 5.75--5.95 & 5.832 & 23 & 295 & 189 & $0.655+0.041(+0.080)$ & 70 & $0.475+0.080(+0.158)$ \\
& 5.95--6.15 & 6.043 & 10 & 145 & 113 & $0.797+0.056(+0.110)$ & 49 & $0.676+0.118(+0.249)$ \\
& 6.15--6.35 & 6.225 & 3  & 45  & 38  & $0.864+0.055(+0.116)$ & 20 & $0.889+0.149(+0.311)$ \\
\hline Ly$\alpha$+Ly$\beta$ & 4.95--5.15 & 5.112 & 7 & 73 & 7 & $0.100+0.054(+0.129)$ & 1 & $0.055+0.086(+0.247)$ \\
& 5.15--5.35 & 5.272 & 22 & 361 & 44 & $\mathbf{0.127+0.022(+0.045)}$ & 5 & $\mathbf{0.055+0.031(+0.075)}$ \\
& 5.35--5.55 & 5.450 & 30 & 735 & 144 & $0.205+0.035(+0.076)$ & 23 & $0.125+0.030(+0.062)$ \\
& 5.55--5.75 & 5.643 & 33 & 754 & 203 & $0.282+0.033(+0.066)$ & 39 & $0.207+0.048(+0.104)$ \\
& 5.75--5.95 & 5.832 & 26 & 441 & 161 & $0.382+0.045(+0.090)$ & 35 & $0.317+0.057(+0.119)$ \\
& 5.95--6.15 & 6.038 & 10 & 173 & 81  & $0.490+0.074(+0.155)$ & 17 & $0.393+0.100(+0.219)$ \\
& 6.15--6.35 & 6.225 & 3  & 46  & 26  & $0.591+0.189(+0.324)$ & 7  & $0.609+0.241(+0.533)$ \\
\hline Ly$\beta$+Ly$\gamma$ & 5.35--5.55 & 5.481 & 8 & 132 & 20 & $0.158+0.076(+0.167)$ & 1 & $\mathbf{0.030+0.048(+0.141)}$ \\
& 5.55--5.75 & 5.654 & 24 & 380 & 88 & $0.242+0.046(+0.097)$ & 9  & $\mathbf{0.095+0.037(+0.085)}$ \\
& 5.75--5.95 & 5.831 & 23 & 293 & 86 & $0.307+0.039(+0.079)$ & 14 & $\mathbf{0.191+0.056(+0.126)}$ \\
& 5.95--6.15 & 6.043 & 10 & 141 & 60 & $0.445+0.082(+0.179)$ & 7  & $\mathbf{0.199+0.087(+0.202)}$ \\
& 6.15--6.35 & 6.225 & 3 & 45 & 26 & $\mathbf{0.604+0.153(+0.267)}$ & 7 & $0.622+0.246(+0.542)$ \\
\hline Ly$\alpha$+Ly$\beta$+Ly$\gamma$ & 5.35--5.55 & 5.481 & 8 & 131 & 16 & $\mathbf{0.131+0.056(+0.117)}$ & 0 & $0.000+0.061(+0.237)$ \\
& 5.55--5.75 & 5.655 & 24 & 377 & 78 & $\mathbf{0.221+0.047(+0.095)}$ & 5  & $0.106+0.058(+0.143)$ \\
& 5.75--5.95 & 5.831 & 23 & 288 & 79 & $\mathbf{0.293+0.039(+0.081)}$ & 11 & $0.306+0.104(+0.237)$ \\
& 5.95--6.15 & 6.043 & 10 & 135 & 50 & $\mathbf{0.396+0.093(+0.205)}$ & 4  & $0.237+0.147(+0.363)$ \\
& 6.15--6.35 & 6.225 & 3 & 45 & 24 & $0.571+0.186(+0.321)$ & 3 & $0.533+0.379(+0.920)$ \\
\hline
\end{tabular}
\caption{Dark pixel statistics from the Ly$\alpha$, Ly$\beta$, and Ly$\gamma$ forests of the E-XQR-30 sample. 
The columns from left to right show the forest transition(s), $z$ range, the average $z$ of the pixels falling within the range $\bar{z}$, the number of quasar lines of sight ($N_{\rm LOS}$) and spectral pixels ($N_{\rm pix}$) covering this range, the number of undetected (${\rm S/N<2}$) pixels $N_{\rm un-det}$, the resulting dark fraction from the threshold method $f_{\rm dark}^{\rm thresh}$ with $+1\sigma$ and $+2\sigma$ uncertainties, the number of negative pixels $N_{\rm neg}$, and the resulting dark fraction from the negative pixel method $f_{\rm dark}^{\rm neg}$ with $+1\sigma$ and $+2\sigma$ uncertainties. Uncertainty ranges for $f_{\rm dark}$ can go above unity due to the re-scaling factors applied to convert the corresponding fractions of un-detected or negative pixels. The fiducial constraints for each method, corresponding to the lowest $2\sigma$ upper limit, are shown in boldface.}
\label{tab:results}
\end{table*}

\subsection{Binned pixel definition}\label{sec:bin}

We follow \citet{McGreer11,McGreer15} and bin the spectra into 3.3\,Mpc segments, corresponding to a velocity scale of $\Delta v\simeq380$\,km/s in our assumed cosmology at $z=5$--$6$. This scale was originally motivated in \citet{McGreer11} for various reasons. Observationally, this binning allows for higher sensitivity to IGM transmission features. From a more theoretical standpoint, Jeans-scale patches of neutral IGM at the end of the reionization process should have a broad, saturated core in addition to damping wing absorption, such that arbitrarily small dark gaps are unlikely to correspond to neutral gas (e.g.~\citealt{Zhu22}). The damping wings associated with such patches likely extend even further than this scale \citep{Zhu24,Spina24}, increasing the potential number of dark pixels. This choice introduces a degree of model dependence to our results -- if pockets of neutral IGM can produce smaller gaps, our binning would miss them. Conversely, if the neutral IGM only ever manifests as larger dark regions, a larger binning scale would exclude the smaller dark regions consistent with the ionized phase of the IGM. We explore the dependence of our results as a function of binning scale in Appendix~\ref{app:bin}.

We select the 3.3\,Mpc segments in fixed redshift ranges by integrating the comoving line element $dl/dz = c/H(z)$ starting from $z=4.75$, the lower redshift end of the lowest redshift Ly$\alpha$ forest bin centered at $z=4.85$. We then average together spectral pixels corresponding to Lyman-series forest redshifts within these 3.3\,Mpc bins with no weighting. The flux error on each binned pixel is then propagated appropriately to compute $\sigma_{\rm bin}$. We remove any binned pixel which has more than 50\% of its spectral pixels masked to avoid mixing together too broad of a range of scales.

To prune lower-quality binned pixels from the data set, we choose to keep pixels with limiting effective optical depth $\tau_{\rm eff, lim}\geq4.0$. While this is slightly less strict than the $\tau_{\rm eff,lim}\geq4.5$ criterion in \citet{McGreer15} and \citet{Jin23}, we find that there is little qualitative difference in our results, unlike the more heterogeneous data sets in those works. Due to the uniformly high ${\rm S/N}$ of the E-XQR-30 data set, the vast majority of binned pixels satisfy this criterion. The only notable exception is the wavelength range of telluric O$_2$ A-band absorption at observed wavelength $\lambda_{\rm obs}\sim7600$\,\AA, visible as the vertical blue stripe at Ly$\alpha$ redshift of $z_{\rm Ly\alpha}\sim 5.25$ in Figure~\ref{fig:taulim}.

\subsection{Dark fraction from the threshold method}\label{sec:thresh}

Two methods were proposed by \citet{Mesinger10} to measure the dark fraction of binned Lyman-series forest pixels: the ``threshold'' method, and the ``negative pixel'' method. The threshold method is the most straightforward and conservative method, whereby binned pixels with ${\rm S/N}$ below some threshold are considered dark. We follow \citet{McGreer11} and identify dark pixels with a ${\rm S/N}<2$ threshold. In the left panel of Figure~\ref{fig:threshpix}, we show the full distribution of detected (${\rm S/N}>2$, blue) and undetected (${\rm S/N}<2$, red) pixels along our quasar sightlines, where the individual quasars have been sorted from highest redshift (top) to lowest redshift (bottom). Pixels within the excluded quasar proximity zones are shown in faded colors. In the right panel of Figure~\ref{fig:threshpix} we show corresponding distributions for the combined forests (Ly$\alpha$+Ly$\beta$, Ly$\beta$+Ly$\gamma$, Ly$\alpha$+Ly$\beta$+Ly$\gamma$), requiring undetected pixels to be undetected in all considered transitions.

For our threshold ${\rm S/N}<2$, we expect that a small fraction ($\simeq2.3\%$) of truly dark regions will have measured flux values that scatter above the threshold assuming Gaussian noise statistics. Thus we slightly adjust the resulting fraction of undetected pixels, $N_{\rm un-det}/N_{\rm pix}$, by a factor of $1.023$ to estimate the true dark fraction. Multiple Lyman-series forests are combined by requiring ${\rm S/N}<2$ for all transitions in the same pixel, with the corresponding correction factors increased proportional to the number of considered forests. In Figure~\ref{fig:dark_thresh}, we show the dark fraction evolution for every Lyman-series forest and combination considered here (see also Table~\ref{tab:results}).

\subsection{Dark fraction from the negative pixel method}\label{sec:neg}

The second method used to measure the dark fraction is the ``negative pixel'' method, which is in principle more sensitive to truly dark pixels. The negative pixel method assumes that the distribution of truly dark pixels is symmetric about zero, such that twice the number of negative pixels can be used as an estimate of the number of truly dark pixels. While the dark fraction derived from this method is less contaminated by pixels with low levels of transmitted flux, and thus should in general be lower, it relies on extremely accurate sky subtraction to avoid strongly biasing the dark fraction. For combinations of forests, truly dark pixels are selected to be negative in all forests simultaneously, requiring additional factors of two for each considered forest (e.g. 1 Ly$\alpha$+Ly$\beta$ negative pixel $=$ 4 dark pixels).

\begin{figure*}
\begin{center}
\resizebox{8.0cm}{!}{\includegraphics[trim={1.0em 1em 1em 1em},clip]{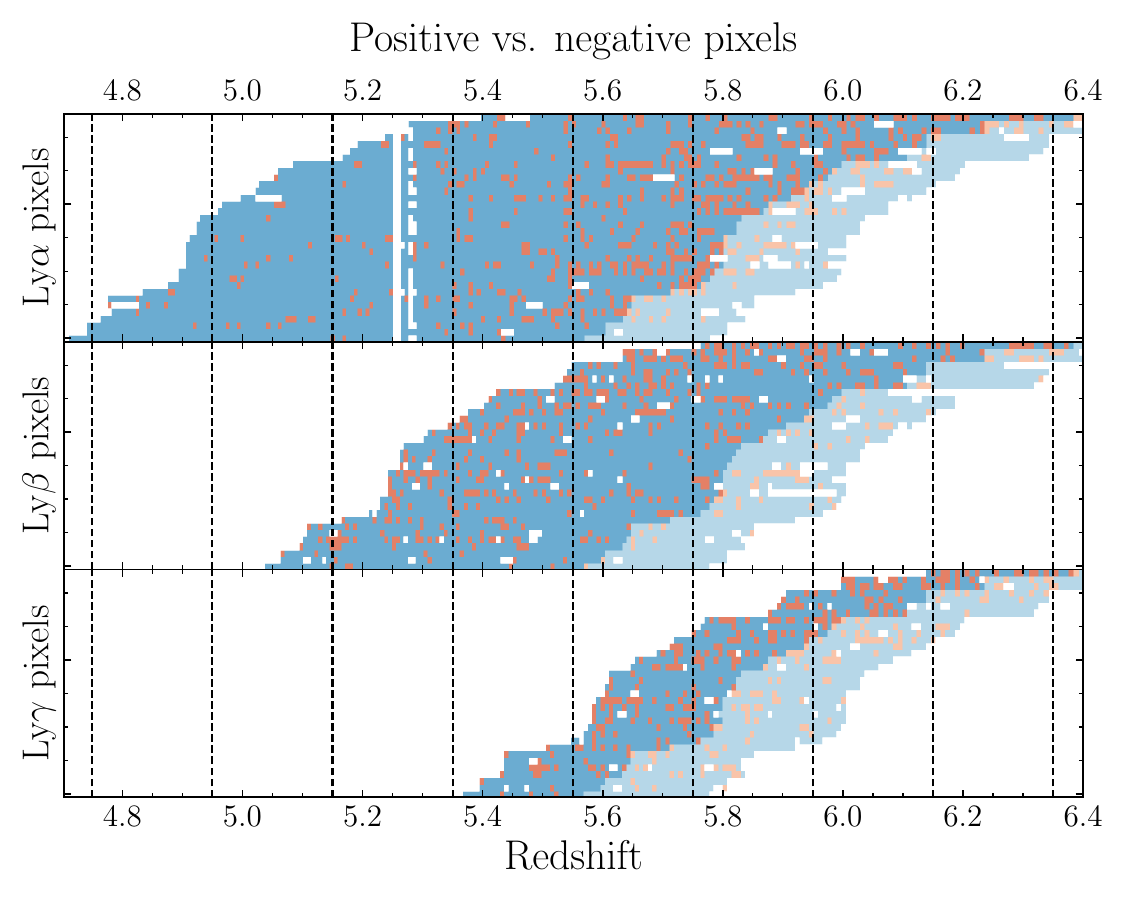}}
\resizebox{8.0cm}{!}{\includegraphics[trim={1.0em 1em 1em 1em},clip]{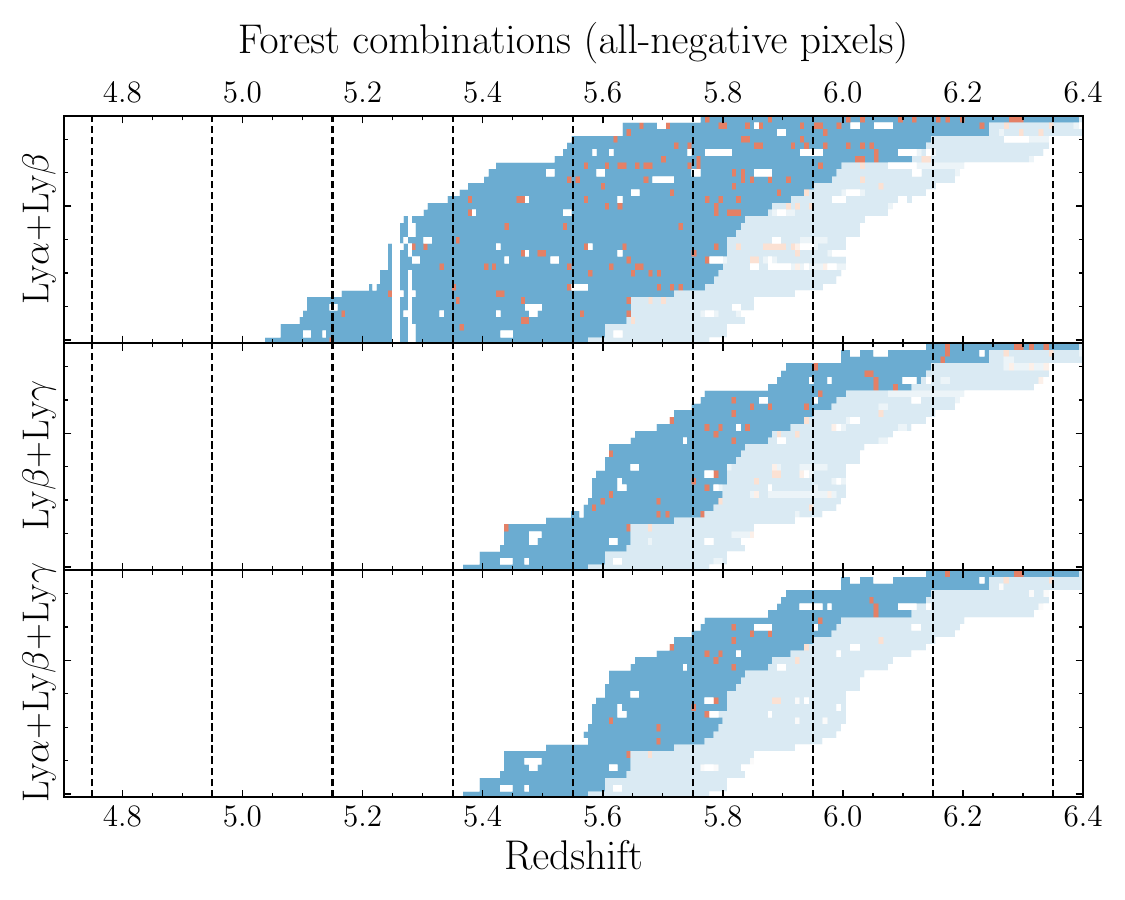}}\\
\end{center}
\vspace{-1.0em}
\caption{Similar to Figure~\ref{fig:threshpix} but for the negative pixel method, with positive pixels in blue and negative pixels in red. }
\label{fig:negpix}
 \end{figure*}

\begin{figure*}
\begin{center}
\resizebox{11cm}{!}{\includegraphics[trim={1.0em 1em 1em 0.5em},clip]{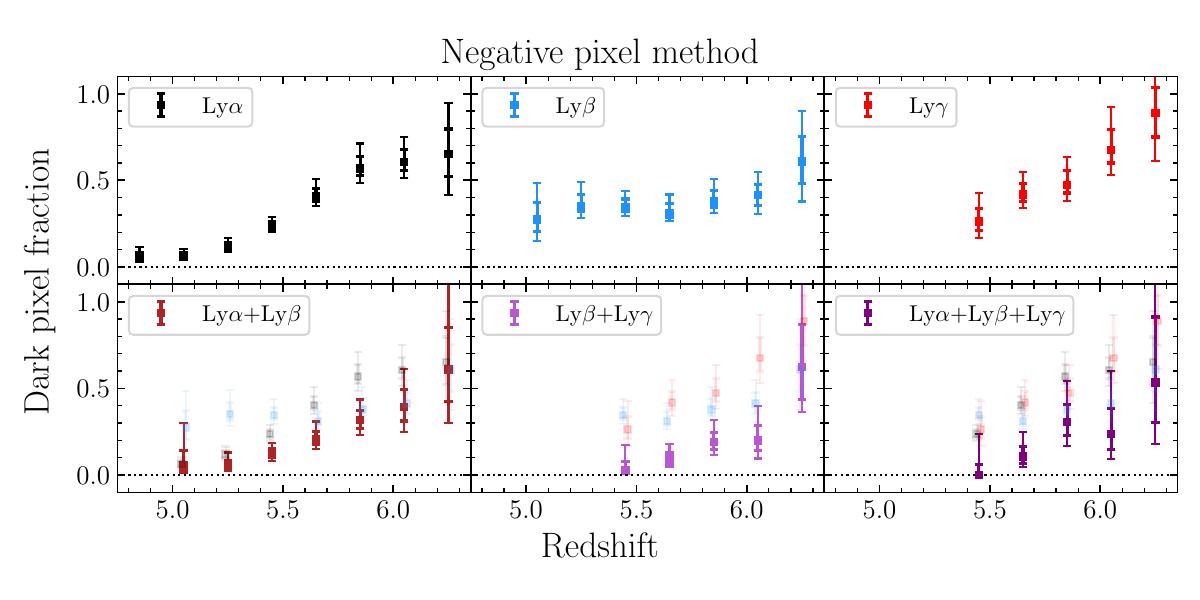}}
\end{center}
\vspace{-2.0em}
\caption{Dark pixel fraction measurements for all forest combinations similar to Figure~\ref{fig:dark_thresh}, but for the negative pixel method.}
\label{fig:dark_negpix}
 \end{figure*}

In the left panel of Figure~\ref{fig:negpix} we show the distribution of positive (blue) and negative (red) pixels in our quasar spectra, similar to Figure~\ref{fig:threshpix}. The fraction of negative pixels increases with redshift as the transmitted flux of the forest drops, and more of the spectrum shows no visible transmission. In the right panel of Figure~\ref{fig:negpix} we show the combined forest distributions, again similar to Figure~\ref{fig:threshpix} but now requiring the pixels to be negative in all considered transitions. In Figure~\ref{fig:dark_negpix}, we show the resulting binned dark pixel fractions for every forest combination, similar to Figure~\ref{fig:dark_thresh}.

 \begin{figure*}
\begin{center}
\resizebox{8.7cm}{!}{\includegraphics[trim={0.5em 1em 1em 1em},clip]{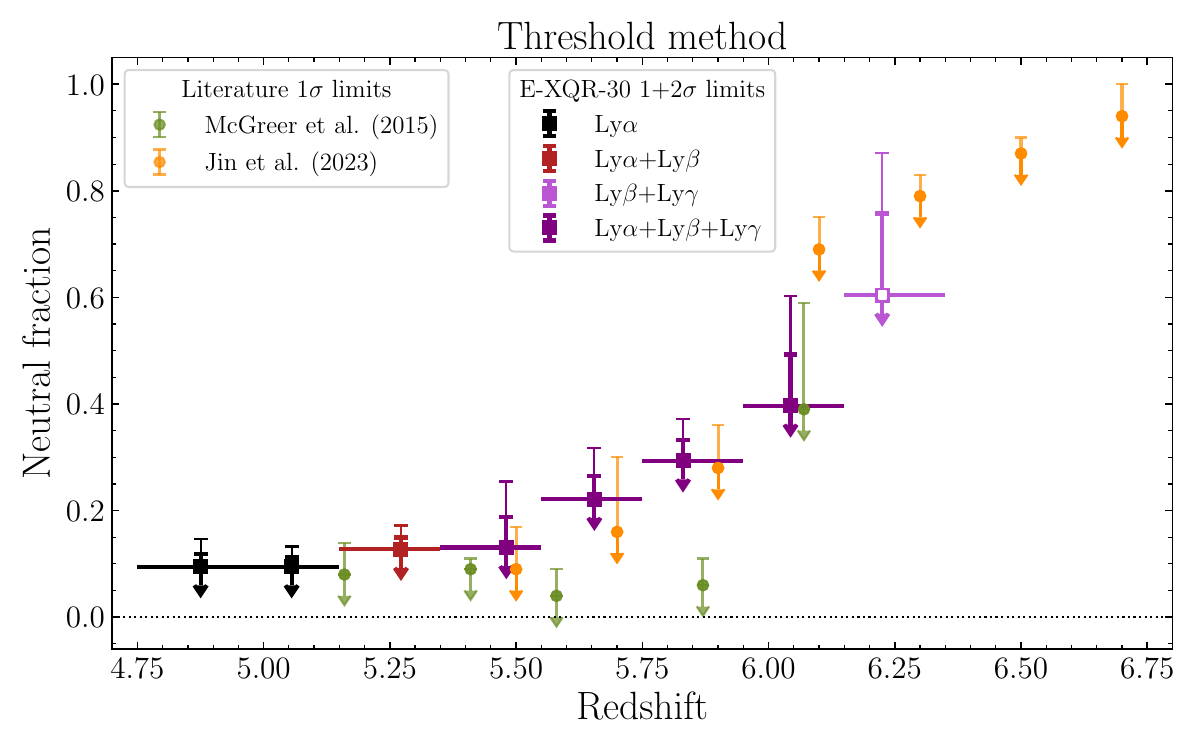}}
\resizebox{8.7cm}{!}{\includegraphics[trim={0.5em 1em 1em 1em},clip]{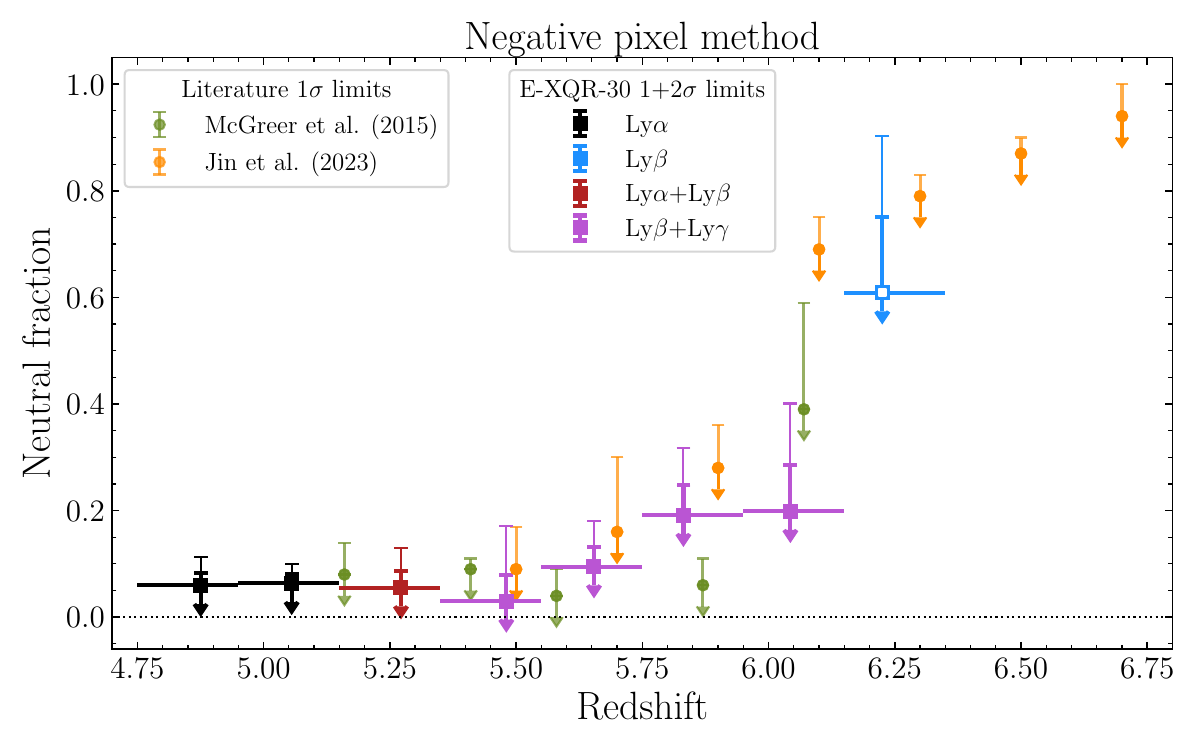}}
\end{center}
\vspace{-1.0em}
\caption{Upper limits ($1\sigma$, thick; $2\sigma$, thin) on the IGM neutral fraction \xhi\ from the dark pixel fractions estimated using the threshold method (left) and negative pixel method (right). In each redshift bin we show the fiducial forest/combination that has the lowest $2\sigma$ upper limit (see text). We compare our constraints to the dark pixel fraction measurements ($1\sigma$ upper limits) from \citet{McGreer15} (green points) and \citet{Jin23} (orange points), which were estimated using the negative pixel method and threshold method, respectively.}
\label{fig:xhi_fiducial}
 \end{figure*}

\subsection{Upper limits on the IGM neutral fraction}\label{sec:indep_const}

As described in Section~\ref{sec:intro}, the dark pixel fractions from the threshold and negative pixel methods represent an upper limit to \xhi. It is thus important that we carefully quantify the uncertainty. Past works have evaluated the uncertainty using the jackknife method applied at the quasar sightline level \citep{McGreer11,McGreer15,Jin23}. Due to our increased sample size, we instead opt for a bootstrap over quasar sightlines, and record both the $1\sigma$ and $2\sigma$ uncertainties from the percentiles of the bootstrap draws. We find that, in general, our bootstrap uncertainties are larger than those determined from jackknife, making our choice a conservative one. 

However, we note that at some level the measurement of the dark fraction represents a binomial process, i.e. we are constraining some fraction of ``successes'' from a discrete number of draws. What separates our problem from being strictly binomial is that there are strong correlations along individual lines of sight (e.g. Gunn-Peterson troughs, \citealt{Becker15,Zhu21,Zhu22}). Nevertheless, these correlations should only serve to \emph{increase} the uncertainty in the dark fraction by reducing the effective number of independent draws. Thus an estimate of the uncertainty of the corresponding binomial process using the number of total draws $n_{\rm tot}$ and the number of ``successes'' (dark pixels) $n_{\rm dark}$ should represent a lower limit on the uncertainty of the dark fraction. In practice, we apply Wilson's score interval \citep{Wilson27} with $z_\alpha=1$ and $2$ to recover the $1\sigma$ and $2\sigma$ upper bounds, respectively. In the negative pixel method, because what is actually measured is the fraction of \emph{negative} pixels, we reduce the number of successes accordingly (i.e. to $n_{\rm neg}$) and recover the dark fraction uncertainty by multiplying the negative fraction by its corresponding conversion factor. Because the binomial uncertainty is a lower limit to the uncertainty in the occurrence rate of dark pixels, when the uncertainty estimated by bootstrap is smaller than the corresponding binomial uncertainty, we replace it with the binomial uncertainty. This replacement almost never occurs in the threshold method, but almost always occurs for the negative pixel method forest combinations where the effective number of pixels is smaller by a factor of four or larger. For example, the binomial uncertainties of the negative pixel method dark fraction of the Ly$\beta$+Ly$\gamma$ forest are larger by a factor of $\sim1.5$ than the bootstrap estimate.

For each dark fraction measurement method we have upper limits to \xhi\ for as many as six different Lyman-series forests and combinations thereof, see Table~\ref{tab:results}. We select the most constraining forest or combination for each method in each redshift bin by choosing the lowest $2\sigma$ upper limit, which is more sensitive to the uncertain tail from small-number statistics than the $1\sigma$ upper limit. In Figure~\ref{fig:xhi_fiducial}, we show our fiducial \xhi\ constraints from the threshold method (left) and the negative pixel method (right) compared to previous measurements by \citet{McGreer15} and \citet{Jin23}. While the threshold method tends to benefit from combining as many forests as possible, the negative pixel method increasingly suffers from small-number statistics, particularly when all three forests are combined. In the threshold method, the most constraining measurements at the most interesting redshifts $z\sim5.5$--$6.0$ come from the full combination of the Ly$\alpha$$+$Ly$\beta$$+$Ly$\gamma$ forests, while in the negative pixel method, the Ly$\beta$$+$Ly$\gamma$ forest combination is the most sensitive. The higher order Lyman series lines are generally more transparent at fixed neutral fraction (e.g.~\citealt{OF05}), permitting weak transmission spikes that tilt the statistics of some Ly$\alpha$-dark ionized regions towards positive flux values. As expected, similar to \citet{McGreer15}, we find that the negative pixel method is the most constraining -- we will consider its corresponding \xhi\ constraints as the fiducial measurements of this work. 

Our fiducial \xhi\ constraints from the negative pixel method are $\langle x_{\rm HI} \rangle \leq \{0.030+0.048,0.095+0.037,0.191+0.056,0.199+0.087\}$ at $\bar{z}=\{5.481,5.654,5.831,6.043\}$ from the Ly$\beta+$Ly$\gamma$ forest combination. In our highest redshift bin, which is only covered by three quasar sightlines, we measure $\langle x_{\rm HI} \rangle \leq 0.609+0.147$ at $\bar{z}=6.225$ -- however, with such poor sightline sampling, we consider this upper limit tentative. 

\section{Model-dependent IGM neutral fraction constraints}\label{sec:dep}

The model-independent methods described in the previous section make the very conservative assumption that any dark regions of the spectrum \emph{could} be neutral IGM, but a substantial fraction of the ionized IGM will be dark at current spectroscopic sensitivities. The key question is exactly what fraction of dark pixels are neutral and truly dark vs. ionized and lying below detection limits. One could perform a fully Bayesian analysis by forward-modeling quasar spectra using 3D ionization and thermal histories computed from large-volume radiation-hydrodynamical simulations of the IGM and comparing them to the observations via a set of summary statistics. Due to the wide dynamic range required to accurately simulate the reionization process while simultaneously resolving the Ly$\alpha$ forest, this has so far only been done for large-scale-averaged Ly$\alpha$ forest opacity with approximate, semi-numerical approaches \citep{Qin21,Qin24}. In this section, we describe a parametric, model-\emph{dependent} method to constrain the dark pixel fraction in the Lyman-series forests. 

\begin{figure*}
\begin{center}
\resizebox{8.0cm}{!}{\includegraphics[trim={0.5em 1em 1em 1em},clip]{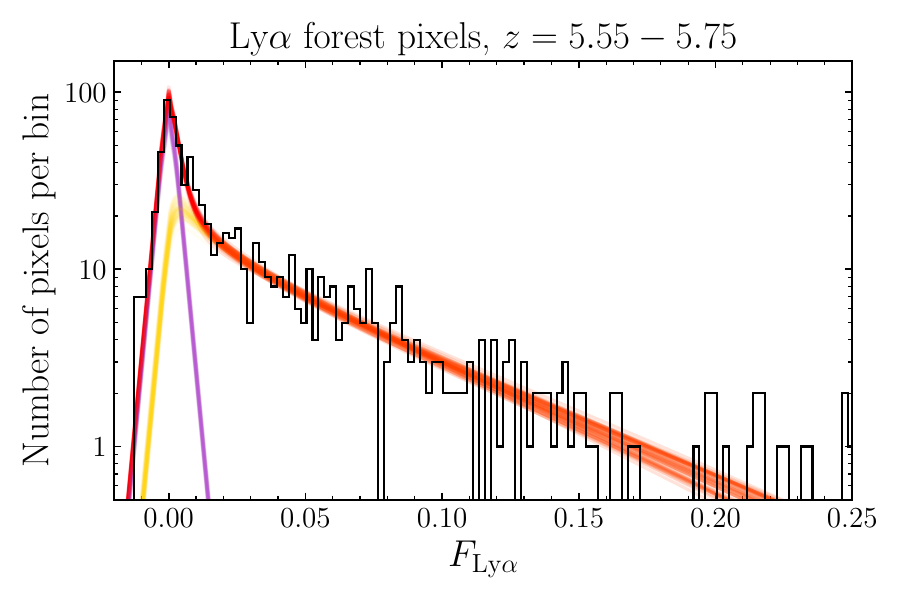}}
\resizebox{8.0cm}{!}{\includegraphics[trim={0.5em 1em 0.5em 0.5em},clip]{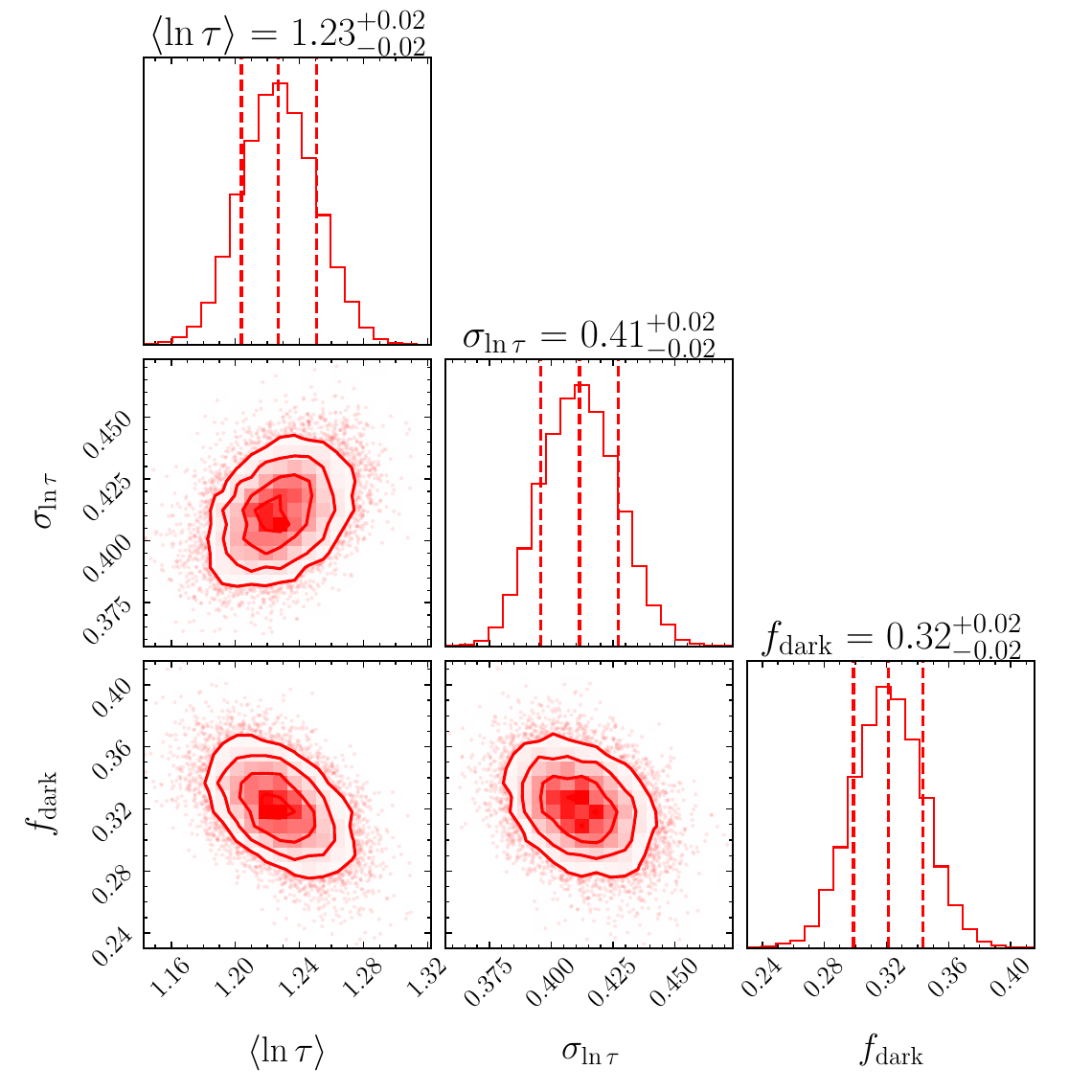}}
\end{center}
\caption{Lognormal mixture fit to the transmitted flux PDF of the Ly$\alpha$ forest at $z=5.55$--$5.75$. Left: The flux PDF data are shown in black, along with 50 red curves corresponding to posterior draws of the fit parameters. The purple and yellow curves correspond to the ``dark'' and ``transparent'' components of the model draws, respectively. Right: Posterior PDF of the lognormal model parameters $\langle \ln{\tau} \rangle$, $\sigma_{\ln{\tau}}$ as well as the dark fraction $f_{\rm dark}$.}
\label{fig:mixfit_demo}
 \end{figure*}

 \begin{figure*}
\begin{center}
\resizebox{8.0cm}{!}{\includegraphics[trim={0.5em 1em 1em 1em},clip]{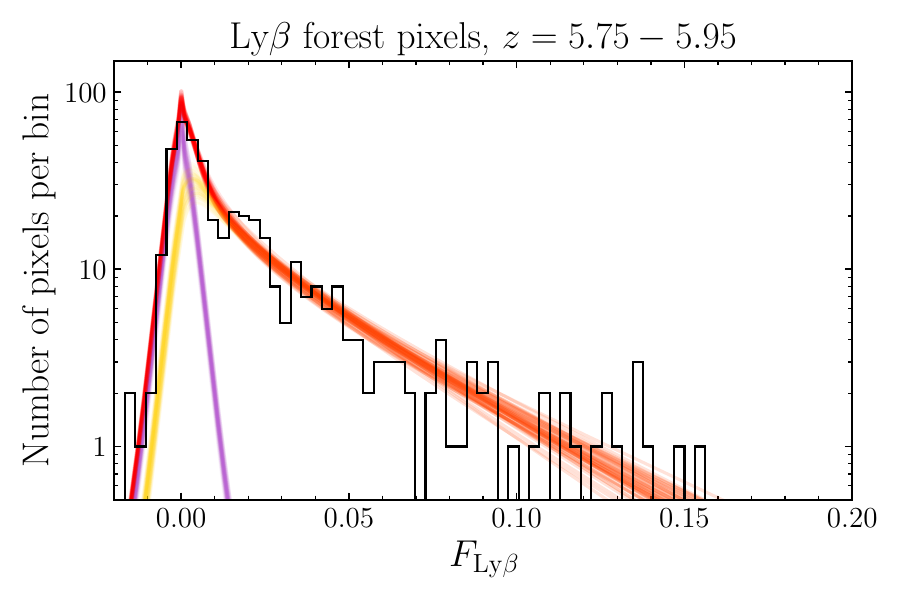}}
\resizebox{8.0cm}{!}{\includegraphics[trim={0.5em 1em 0.5em 0.5em},clip]{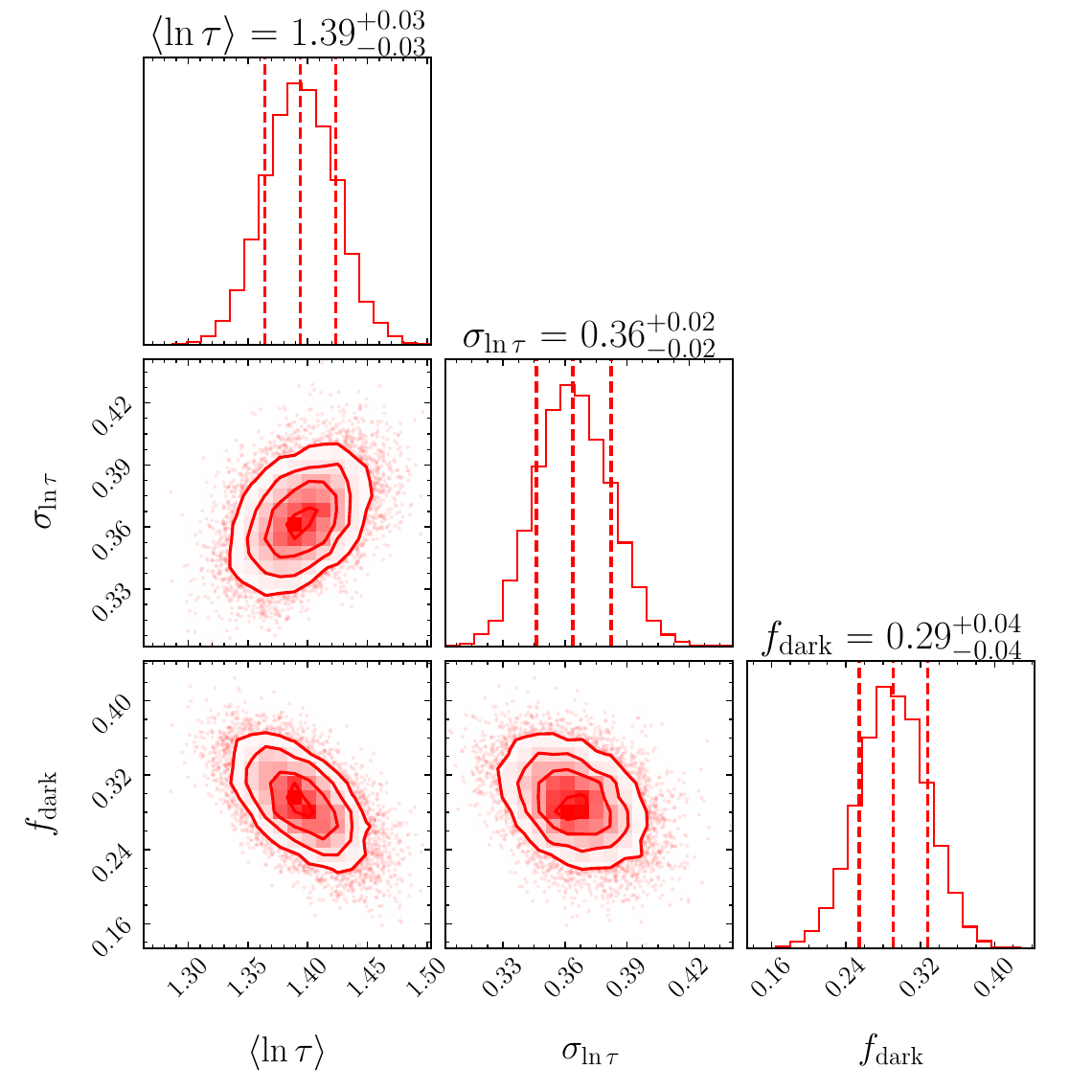}}
\end{center}
\caption{Lognormal mixture fit to the transmitted flux PDF of the Ly$\beta$ forest at $z=5.75$--$5.95$, similar to Figure~\ref{fig:mixfit_demo}.}
\label{fig:mixfit_demo2}
\end{figure*}

\begin{figure*}
\begin{center}
\resizebox{11cm}{!}{\includegraphics[trim={1.0em 1em 1em 0.5em},clip]{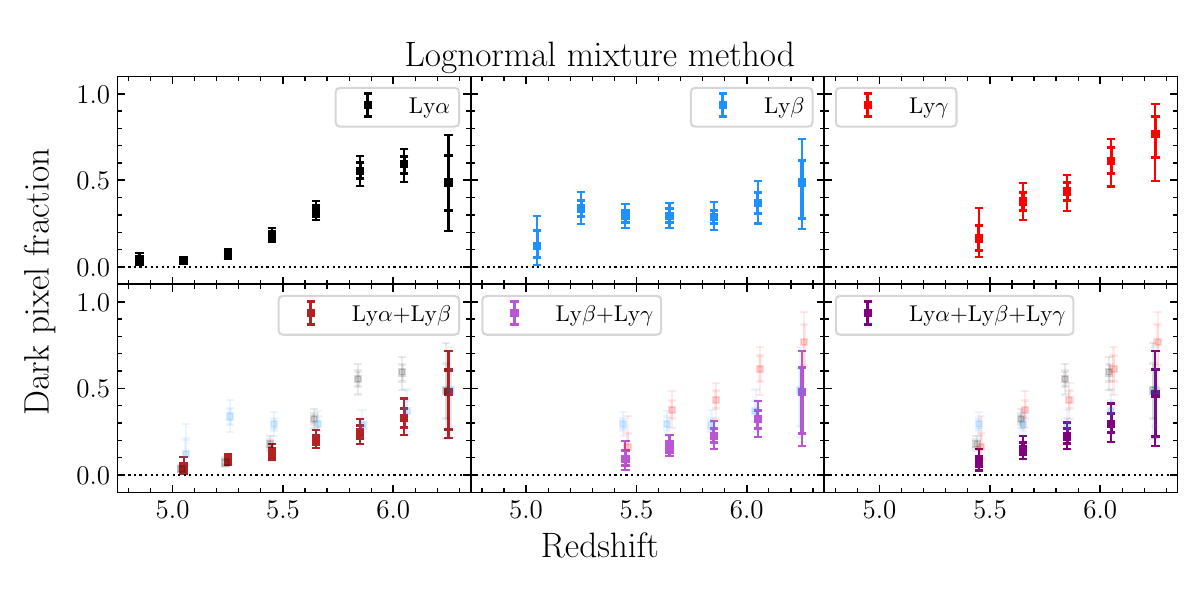}}
\end{center}
\vspace{-2.0em}
\caption{Upper limits on \xhi\ from all forest combinations similar to Figure~\ref{fig:dark_thresh} but for the model-dependent lognormal mixture method.}
\label{fig:dark_mixfit}
\end{figure*}

It is well established that the cosmological density distribution is approximately lognormal on large scales \citep{CJ91}. The Ly$\alpha$ (and Lyman-series) optical depth is proportional to density to some power via the fluctuating Gunn-Peterson approximation (FGPA; \citealt{Weinberg97}), thus it is reasonable to expect that the Ly$\alpha$ optical depth is also (at least roughly) lognormally distributed. To a good approximation this is indeed the case in the observed high-resolution Ly$\alpha$ forest at all redshifts \citep{Becker07}, as well as on larger binning scales in hydrodynamical simulations of the IGM at $z\sim6$ (cf. the appendix of \citealt{Davies20a}, see also \citealt{Maity25}). More generally, \citet{Becker07} argue that a lognormal optical depth distribution could arise due to random sampling of the multiplicative terms in the FGPA (i.e. the gas temperature and hydrogen ionization rate).

We can then construct a simple model for the distribution of forest flux by considering the spectrum as a random mixture of ``dark'' pixels, whose transmitted flux is drawn from a Gaussian centered at zero, and ``transparent'' pixels, whose transmitted flux is drawn from the convolution of a lognormal optical depth distribution with the noise in each pixel. That is, we can write the probability distribution function (PDF) for the flux $F_i$ in a given pixel $i$ as
\begin{equation}
    P(F_i|\sigma_i,f_{\rm dark}) =  f_{\rm dark} P(F_i|\sigma_i,{\rm dark})+\,(1-f_{\rm dark})\, P(F_i|\sigma_i,\!\sim\!{\rm dark}),
\end{equation}
where $\sigma_i$ is the noise in pixel $i$, $f_{\rm dark}$ is the fraction of truly dark pixels, $P(F_i|\sigma_i,{\rm dark})$ is the flux PDF of dark pixels, and $P(F_i|\sigma_i,\!\sim\!{\rm dark})$ is the flux PDF of transparent (not-dark) pixels. The flux PDF of dark pixels is simply a normal distribution $\mathcal{N}$ centered on zero with width $\sigma_i$,
\begin{equation}
    P(F|\sigma_i,{\rm dark}) = \mathcal{N}(F|0,\sigma_i),
\end{equation}
while the transparent flux PDF is the convolution of the noise with the intrinsic flux PDF $P(F|\sim\!{\rm dark})$,
\begin{equation}
    P(F_i|\sigma_i,\sim\!{\rm dark}) = (P(F|\sim\!{\rm dark}) * \mathcal{N}(F|0,\sigma_i))(F_i),
\end{equation}
where we assume that $P(F|\sim\!{\rm dark})$ follows a lognormal distribution in (effective) optical depth. This lognormal optical depth distribution maps to flux space as (e.g.~\citealt{Becker07})
\begin{equation}
\begin{aligned}
    P(F|\!\sim\!{\rm dark}) & = \\ 
    & \frac{1}{(-\ln{F})F\sigma_{\ln{\tau}}\sqrt{2\pi}}\exp{\left(-0.5\frac{(\ln{(-\ln{F})}-\langle\ln{\tau}\rangle)^2}{\sigma_{\ln{\tau}}^2}\right)}
\end{aligned}
\end{equation}
where $\langle\ln{\tau}\rangle$ and $\sigma_{\ln{\tau}}$ are the mean and width of the lognormally distributed optical depth. This mixture model can also be used to estimate the probability that any given pixel is drawn from the dark or transparent distributions,
\begin{equation}\label{eqn:pdark}
    P({\rm dark}|F_i) = \frac{f_{\rm dark}P(F_i|{\rm dark})}{f_{\rm dark}P(F_i|{\rm dark})+(1-f_{\rm dark})P(F_i|\sim\!{\rm dark})}.
\end{equation}
This method can be considered similar to the threshold method, but with the addition of a model-based prediction for the number of pixels with ${\rm S/N}<2$ that are drawn from the same distribution as the transparent pixels with ${\rm S/N}>2$. Thus we expect tighter upper limits as compared to the threshold method, which treats all pixels with ${\rm S/N}<2$ as dark.

Note that we use the $\sigma_i$ corresponding to each data pixel without accounting for correlations with the flux $F_i$, i.e. we assume that the noise is generally sky background-limited. Highly-transmissive regions in the brightest quasar in our sample violate this assumption, but the crucial region of the PDF for discriminating between dark and transparent pixels is close to $F=0$, where the background-limited assumption should hold. One possible limitation is that the width of the dark PDF will be overestimated due to ``excess'' Poisson noise present in the distribution of $\sigma_i$ within flux spikes, however, we do not see evidence for overestimation of the width of the dark part of the flux PDF when comparing to the actual data.

We have also neglected to include any consideration of the uncertainty in the quasar continuum. For the simple power-law model of the continuum we have adopted, the relative error in the continuum is likely to be $\simeq15\%$ along with a similar degree of bias \citep{Bosman21}. The effect of continuum error is multiplicative, similar to the physical processes underlying the lognormal optical depth distribution model, so it is likely that the bias and scatter of the true continuum relative to estimates will be absorbed into the $\langle \ln{\tau} \rangle$ and $\sigma_{\ln{\tau}}$ parameters, respectively. 

In the following, we will also apply this model to the Ly$\beta$ and Ly$\gamma$ forests, but note that this may violate the above justifications of lognormality, as the Lyman-series forests are more generally described by a \emph{sum} of optical depths including all lower series lines. In practice, however, we find that the distribution of transparent fluxes is nevertheless well-described by our simple model.

Before proceeding, we note an important caveat to this approach. While the Ly$\alpha$ forest of a uniformly-ionized IGM may be well-described by a lognormal optical depth distribution, there may be a complicated (i.e. non-lognormal) distribution of UV background intensities on large scales \citep{MF09,DF16}. If there is a tail to low UV background intensity, e.g. due to shadowing effects by neutral islands \citep{Kulkarni19,ND20,Choudhury21}, these regions will be degenerate with the truly dark pixels of the neutral IGM. Thus we consider the dark fraction obtained with this method to still represent an upper limit to \xhi.

We now have a three-parameter model that can be fit to each redshift bin of the Ly$\alpha$, Ly$\beta$, and Ly$\gamma$ forests. In Figure~\ref{fig:mixfit_demo} and Figure~\ref{fig:mixfit_demo2} we show examples of this fitting procedure applied to the Ly$\alpha$ forest at $z=5.55$--$5.75$ and the Ly$\beta$ forest at $z=5.75$--$5.95$, respectively, along with corner plots showing parameter uncertainties derived via Markov-chain Monte Carlo (MCMC) sampling. The uncertainty on the dark fraction derived from the MCMC posterior is underestimated as our model does not account for line-of-sight correlations due to cosmic variance; more precisely, the expression given in equation~(\ref{eqn:pdark}) does not include information from neighboring pixels, while real neutral regions can be contiguous over scales larger than one 3.3\,Mpc pixel (e.g.~\citealt{Becker15}). To approximately account for this additional uncertainty, we bootstrap on quasar sightlines as in Section~\ref{sec:indep_const} and re-measure $f_{\rm dark}$ by computing the average $P({\rm dark})$ inferred by applying equation~(\ref{eqn:pdark}) to each bootstrap sample, while simultaneously drawing model parameters from the MCMC samples derived from the true data set. The resulting bootstrap uncertainties are typically a factor of a few larger than those derived directly from the posterior PDF (e.g. the right panels of Figure~\ref{fig:mixfit_demo} and Figure~\ref{fig:mixfit_demo2}).

Deriving quantitative \xhi\ constraints from forest combinations in this framework is exceedingly complicated -- in principle, one would need to carefully model the expected flux correlations between each forest including dilution by foregrounds. We instead opt for a simple approach motivated by the threshold method. We measure $f_{\rm dark}$ from the average of $P({\rm dark})$ similar to the above for a fiducial transition, but for each pixel $i$, if the $P({\rm dark}|F_i)$ is smaller than some threshold for \emph{any} of the considered forests, we set $P({\rm dark})$ to that small value. Uncertainties are again computed via bootstrap as described above. We choose a probability threshold of 1\%, but find that our results are insensitive to this exact choice.

\begin{figure}
\begin{center}
\resizebox{8.7cm}{!}{\includegraphics[trim={1.0em 1em 1em 0.5em},clip]{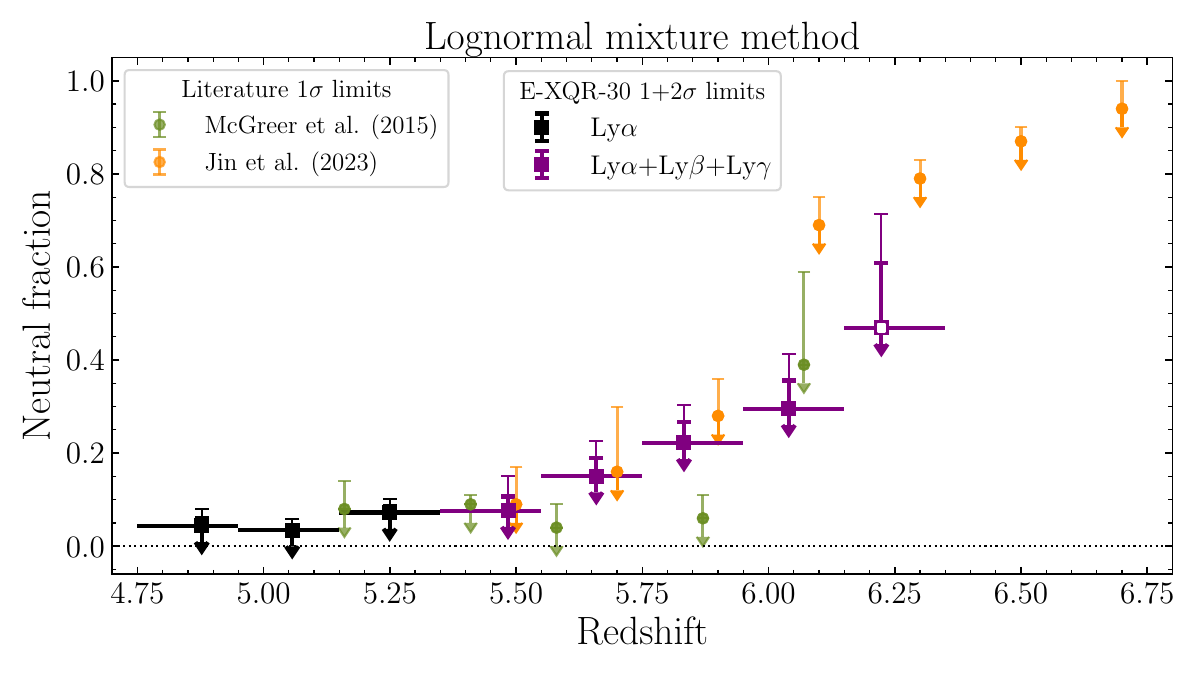}}
\end{center}
\caption{Similar to Figure~\ref{fig:xhi_fiducial} but now for the upper limits on \xhi\ from the model-dependent lognormal mixture method.}
\label{fig:xhi_mixfit}
\end{figure}

In Figure~~\ref{fig:dark_mixfit}, we show the corresponding dark fractions for each forest and combination thereof. Comparison to Figure~\ref{fig:dark_thresh} and Figure~\ref{fig:dark_negpix} suggests that the inferred dark fractions behave similarly with redshift. Choosing the best constraints via the $2\sigma$ upper limits as before, in Figure~~\ref{fig:xhi_mixfit} we show our most constraining model-dependent upper limits on \xhi. From the optimal combination of the Ly$\alpha$+Ly$\beta$+Ly$\gamma$ forests, we find $\langle x_{\rm HI} \rangle \leq \{0.151+0.039,0.222+0.044,0.295+0.061\}$ at $\bar{z}=\{5.654,5.832,6.041\}$. Compared to the model-independent approaches in the previous section, these limits fall below the threshold method, as expected, but they are still higher than the negative pixel method. This trend is likely due to the different assumptions each method makes regarding the pixel statistics -- the threshold method assumes that dark pixels are undetected, the lognormal mixture method assumes that dark pixels are Gaussian-distributed about zero, and the negative pixel method assumes that dark pixels are symmetrically distributed about zero.

Finally, we note that this method should be considered a proof of concept -- we will treat the model-independent measurements from the previous section as our fiducial \xhi\ upper limits. In future work we will explore alternative model-dependent methods built from physical models of IGM transmission, rather than the toy parametric approach considered here.

\section{Discussion \& Conclusion}

\begin{figure}
\begin{center}
\resizebox{8.5cm}{!}{\includegraphics[trim={1.0em 1em 1em 0.5em},clip]{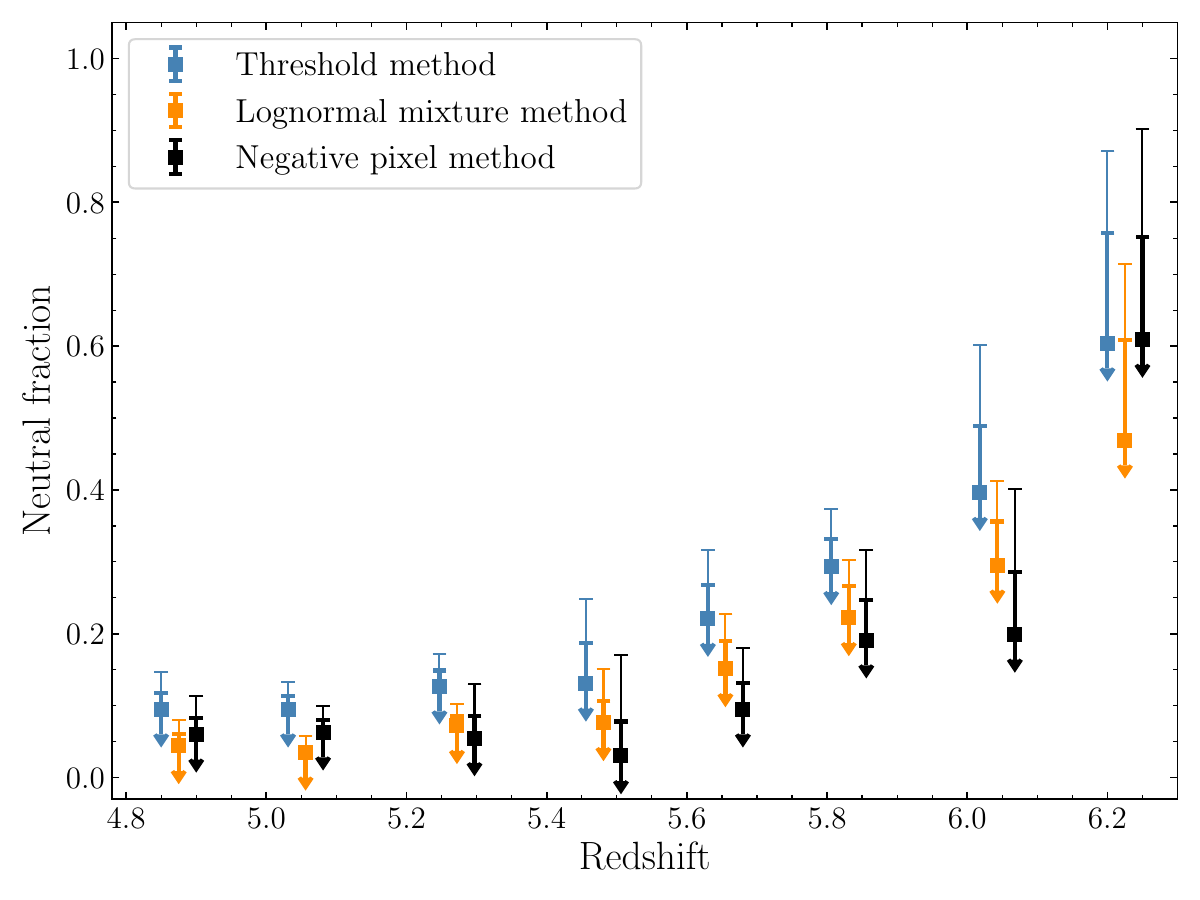}}
\end{center}
\caption{Comparison between our fiducial \xhi\ upper limits from the threshold method (blue), the negative pixel method (black), and the lognormal mixture method (orange). The central redshifts have been offset slightly for clarity.}
\label{fig:compare}
\end{figure}

We have used the exquisite X-Shooter spectra from E-XQR-30 to update the constraints on the dark pixel fraction of the IGM at $z\sim5$--$6$. We carefully clean the spectra of high-variance pixels, and remove spectra which show signs of inaccurate sky subtraction or very strong BAL absorption. We push the constraints further down the Lyman series, and measure dark pixel fractions from the Ly$\alpha$, Ly$\beta$, and Ly$\gamma$ forests. By extending the Ly$\beta$ and Ly$\gamma$ forests to wavelengths shortward of their nominal end, we dramatically increase their corresponding path length. Similar to \citet{McGreer11,McGreer15} and \citet{Jin23}, we place model-independent upper limits on \xhi\ at $z\leq6$ using the threshold and negative pixel methods but with greatly improved statistics. Our data are generally more sensitive both due to the high signal-to-noise ratio of the spectra and the introduction of the Ly$\gamma$ forest. We also explored constraining the dark fraction using a simple parametric model-dependent method that treats the Lyman-series forests as a mixture of dark and transparent regions, finding similar results to the model-independent methods. In Figure~\ref{fig:compare}, we compare the limits estimated from all three methods, with fiducial constraints compiled in Table~\ref{tab:results_fid}.

\begin{table*}\centering
\begin{tabular}{c c c c}
$\bar{z}$ & $\langle x_{\rm HI}\rangle^{\rm thresh}+1\sigma(+2\sigma)$ & $\langle x_{\rm HI}\rangle^{\rm neg}+1\sigma(+2\sigma)$ & $\langle x_{\rm HI}\rangle^{\rm mixture}+1\sigma(+2\sigma)$\\
\hline \hline
4.875 & $0.095+0.023(0.052)$ & $\mathbf{0.060+0.023(0.053)}$ & $0.044+0.017(0.036)$ \\
5.056 & $0.095+0.018(0.038)$ & $\mathbf{0.063+0.017(0.036)}$ & $0.035+0.001(0.023)$ \\
5.272 & $0.127+0.022(0.045)$ & $\mathbf{0.055+0.031(0.075)}$ & $0.072+0.014(0.030)$ \\
5.481 & $0.131+0.056(0.117)$ & $\mathbf{0.030+0.048(0.141)}$ & $0.076+0.031(0.075)$ \\
5.654 & $0.221+0.047(0.095)$ & $\mathbf{0.095+0.037(0.085)}$ & $0.151+0.039(0.076)$ \\
5.831 & $0.293+0.039(0.081)$ & $\mathbf{0.191+0.056(0.126)}$ & $0.222+0.044(0.081)$ \\
6.043 & $0.396+0.093(0.205)$ & $\mathbf{0.199+0.087(0.202)}$ & $0.295+0.061(0.118)$ \\
6.225 & $0.604+0.153(0.267)$ & $\mathbf{0.609+0.143(0.293)}$ & $0.469+0.139(0.245)$ \\
\hline
\end{tabular}
\caption{Upper limits on the IGM neutral fraction \xhi\ from this work using the threshold method, the negative pixel method, and the lognormal mixture method, from left to right. Each measurement is shown as the central value of the corresponding dark fraction along with its $+1\sigma$ ($+2\sigma$) uncertainty. We consider the results from the negative pixel method, highlighted in bold, as our fiducial values.}
\label{tab:results_fid}
\end{table*}

Even with the extended coverage down to rest-frame $\lambda=915$\,\AA, the path length available for higher order Lyman-series lines is extremely limited due to our broad proximity zone exclusion. In fact, only the next higher order line, Ly$\delta$, has substantially non-zero path length, but the number of available pixels is smaller than Ly$\gamma$ by more than a factor of three (see Figure~\ref{fig:npix}). We also investigated the potential for Ly$\delta$ to improve the constraining power of the dark pixel fraction measurements; while the measured dark fractions are competitive, the overall lack of pixels limits its statistical power (see Appendix~\ref{app:delta} for details).

Despite our substantially larger and more sensitive data set, our dark fraction at $z\sim5.8$ ($f_{\rm dark}=0.19$) is nevertheless much higher than the one at $z\sim5.9$ ($f_{\rm dark}=0.06$) reported by \citet{McGreer15}. In fact, if we consider the Ly$\alpha$+Ly$\beta$ forest in the same redshift range $z=5.77$--$5.97$, we recover an even higher dark fraction of $f_{\rm dark}\simeq0.36$. There are a few possible sources for this discrepancy. One could be a limitation of the relatively small number of quasar sightlines in the \citet{McGreer15} sample (6 vs. our 26), as there is considerable cosmic variance between different lines of sight of the Ly$\alpha$ forest (e.g. \citealt{Bosman22}). It is also possible that the $\Delta z=0.1$ proximity zone exclusion in \citet{McGreer15} was too permissive, artificially increasing the path length of transparent forest used in their analysis. We also note that the upper $1\sigma$ uncertainty of $0.05$ quoted by \citet{McGreer15} is about $2/3$ of the corresponding binomial uncertainty on the dark fraction for 1 Ly$\alpha$+Ly$\beta$ negative pixel out of 76 total pixels, $0.076$. 

Another potential source of the discrepancy is systematic differences in data reduction, as the data set compiled for the E-XQR-30 sample includes several of the same quasar spectra used by \citet{McGreer15}: SDSS J0836$+$0054, ULAS J0148$+$0600, SDSS J1306$+$0356, SDSS J0818$+$1722, CFHQS J1509$-$1749, ULAS J1319$+$0950, and SDSS J1030$+$0524. For example, \citet{McGreer15} report that the giant Gunn-Peterson trough in the quasar ULAS~J0148$+$0600 contains only one 3.3\,Mpc pixel which is negative in both the Ly$\alpha$ and Ly$\beta$ forests; in our reduction of the same data, we instead find four, three of which would fall in the $z\sim5.6$ bin of \citet{McGreer15}. The addition of these three coincident-negative pixels from using our reduction of this one spectrum would have increased their dark fraction estimate in that bin from $0.04$ to $0.13$, with substantial additional error from cosmic variance.  The negative pixel method relies on extremely accurate sky subtraction -- as the signal-to-noise of the data increases, any subtle systematic error that does not average down by combining multiple exposures will have more and more influence on the pixel statistics in dark regions. As discussed in Section~\ref{sec:data}, we have made the conservative choice to remove spectra which show signs of a bias in pixel sign (or ``parity'', see Appendix~\ref{app:parity}) rather than attempt to correct them post hoc (as in \citealt{Jin23}, although they ultimately report constraints only from the threshold method).

\begin{figure}
\begin{center}
\resizebox{8.5cm}{!}{\includegraphics[trim={1.0em 1em 1em 0.5em},clip]{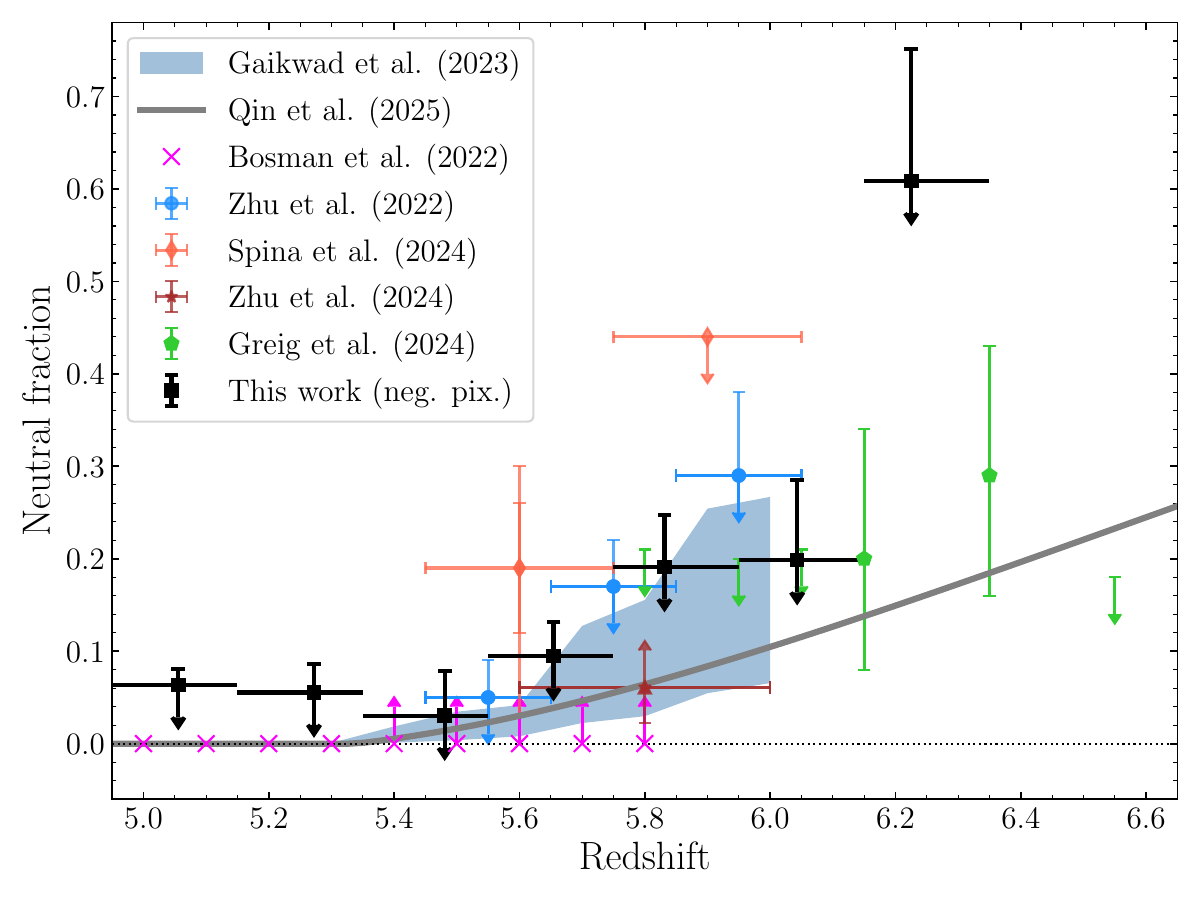}}
\end{center}
\caption{Summary of IGM neutral fraction constraints resulting from other analyses based on the XQR-30 survey compared to the fiducial negative pixel method results in this work (black points). Constraints from the Ly$\alpha$ forest mean flux \citep{Bosman22} are shown as the pink crosses, the Ly$\beta$ dark gap constraints from \citet{Zhu22} are shown as blue points, the Ly$\alpha$ forest damping wing constraints from \citet{Spina24} and \citet{Zhu24} are shown as the orange and brown points, respectively, and the Ly$\alpha$ opacity fluctuations constraints from \citet{Gaikwad23} are shown as a blue swathe. The grey curve shows the maximum a posteriori model from \citet{Qin24}, fit to the Ly$\alpha$ opacity fluctuations. The green points show the (binned) constraints from the quasar damping wing analysis of \citet{Greig24}.}
\label{fig:xhi_lit}
\end{figure}

Our fiducial upper limits on the IGM neutral fraction \xhi\ leave substantial room for incomplete reionization at $z\leq6$, although our limit at $z\simeq6$ more restrictive than other recent works \citep{Zhu22,Jin23}. Compared to the apparently more stringent upper limit from \citet{McGreer15}, our more robust neutral fraction limit of $\langle x_{\rm HI} \rangle \leq 0.2$--$0.3$ at $z\simeq5.9$ has an easier time reconciling the ionizing photon budget and the short mean free path of ionizing photons \citep{Davies21,Cain21}, and is more in line with recent works which tune reionization simulations to match the large-scale fluctuations of the Ly$\alpha$ forest \citep{Kulkarni19,Keating19,Cain23,Cain25jwst,Qin24,Asthana25}. In Figure~\ref{fig:xhi_lit}, we compare our fiducial \xhi\ upper limits to other constraints derived from Ly$\alpha$ transmission in the XQR-30 survey. All of the measurements tell a consistent story of a gradual, late ending to the reionization epoch at $5.3 \lesssim z \lesssim 6.2$; possibly the result of a ``soft landing'' due to recombinations in the IGM \citep{Qin24}, but other alternatives exist such as a rapid fall in the ionizing photon production rate from radiative feedback on low-mass galaxies \citep{Ocvirk21} or an increase in the absorbing cross-section of galactic halos due to outflows \citep{Cain24}.

The model-independent constraints on \xhi\ presented in this work are limited by the fraction of the Ly$\alpha$ and other Lyman-series forests with detectable transmission at high redshift, thus at some point the limits will saturate. At $z\lesssim6$, the limits are unlikely to improve without a substantial increase in sample size at comparable (or better) depth. We do not see a significant trend of the dark fraction with increasing $\tau_{\rm eff, lim}$ within our data set, and reducing the uncertainty of the dark fraction at $z\simeq6$ in the Ly$\beta$$+$Ly$\gamma$ forest by a factor of two would require a $\sim3$--$4$ times larger sample of similar quality. At higher redshifts there is likely room for improvement even with existing data -- careful re-reduction of the three otherwise-uncontaminated $z\sim6.5$ quasars with zero-level drifts would already more than double the path length in our highest redshift bin. In the more distant future, deep spectroscopy of fainter quasar samples (e.g. with ELT/ANDES, \citealt{D'Odorico24}) may allow for smaller proximity zone exclusion, which would increase the spectral coverage of more sensitive higher order Lyman series lines. Ultimately, however, the existing Ly$\alpha$ and other Lyman-series forest data likely already hold tremendous constraining power beyond the \xhi\ upper limits reported here (see \citealt{Qin24}), and await a robust model for IGM transmission on all observed scales during the epoch of reionization.

\section*{Acknowledgements}

We thank Morgan Fouesneau and the MPIA Data Science Department for helpful discussions that led to the lognormal mixture model approach. We also thank Xiangyu Jin for comments on a draft of this manuscript. Part of this work was supported by the German \emph{Deut\-sche For\-schungs\-ge\-mein\-schaft, DFG\/} project number Ts~17/2--1. VD acknowledges financial support from the Bando Ricerca Fondamentale INAF 2022 Large Grant "XQR-30". HC thanks the support by the Natural Sciences and Engineering Research Council of Canada (NSERC), funding reference \#RGPIN-2025-04798 and \#DGECR-2025-00136, and by the University of Alberta, Augustana Campus. MGH has been supported by STFC consolidated grants ST/N000927/1 and ST/S000623/1. FW acknowledges support from NSF award AST-2513040. 

This paper is based on the following ESO observing programmes: XSHOOTER 60.A-9024, 084.A-0360, 084.A-0390, 084.A-0550, 085.A-0299, 086.A-0162, 086.A-0574, 087.A-0607, 088.A-0897, 091.C-0934, 294.A-5031, 096.A-0095, 096.A-0418, 097.B-1070, 098.B-0537, 0100.A-0625, 0101.B-0272, 0102.A-0154, 0102.A-0478, 1103.A-0817.

\section*{Data Availability}

Reduced spectroscopic data from E-XQR-30 is available at \href{https://github.com/XQR-30}{this Github repository}. Any other data underlying this article will be shared on reasonable request to the corresponding author.



\bibliographystyle{mnras}

 \newcommand{\noop}[1]{}



\appendix

\section{Data inspection via ``parity'' tests}\label{app:parity}

\begin{figure*}
\begin{center}
\resizebox{16cm}{!}{\includegraphics[trim={1.0em 1em 1em 0.5em},clip]{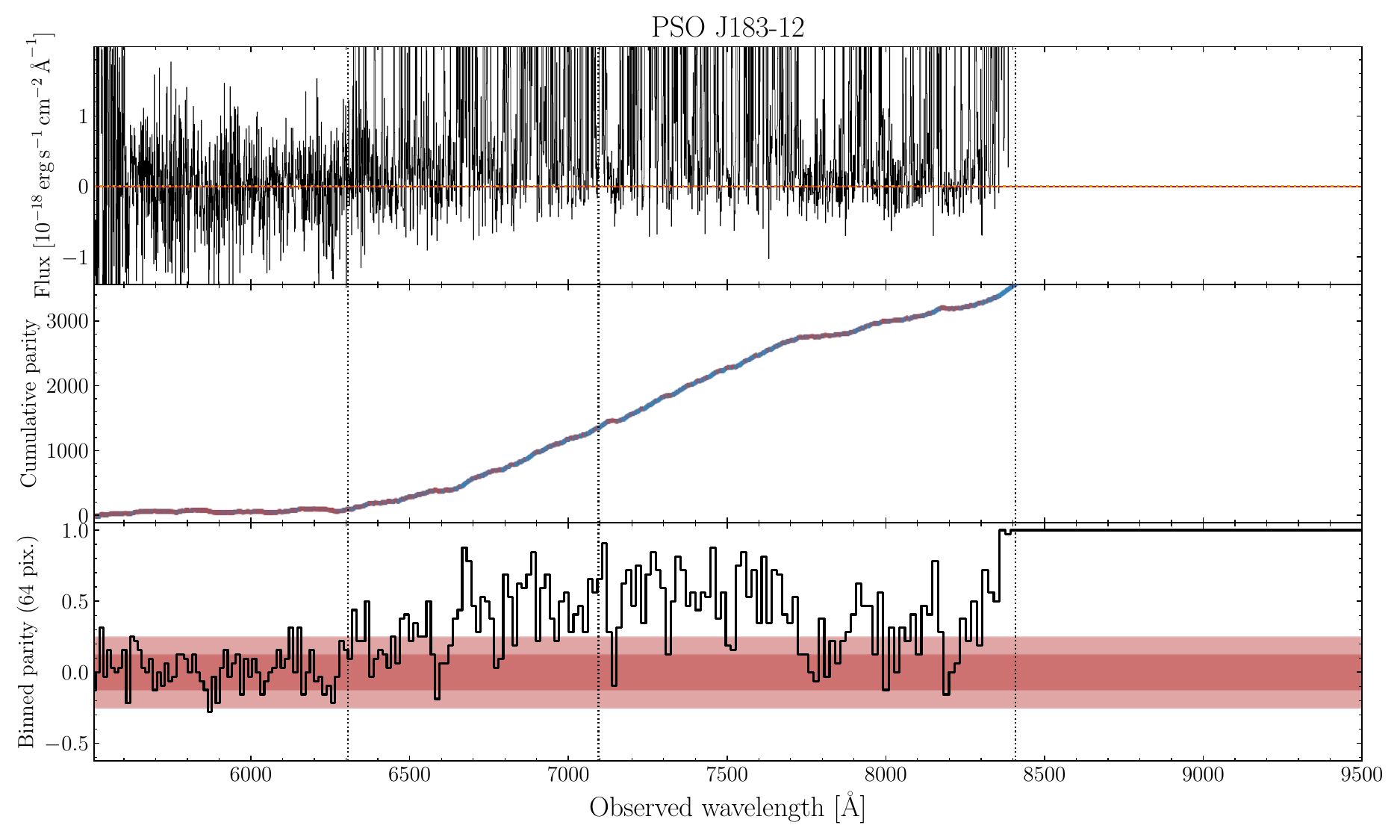}} \\
\resizebox{16cm}{!}{\includegraphics[trim={1.0em 1em 1em 0.5em},clip]{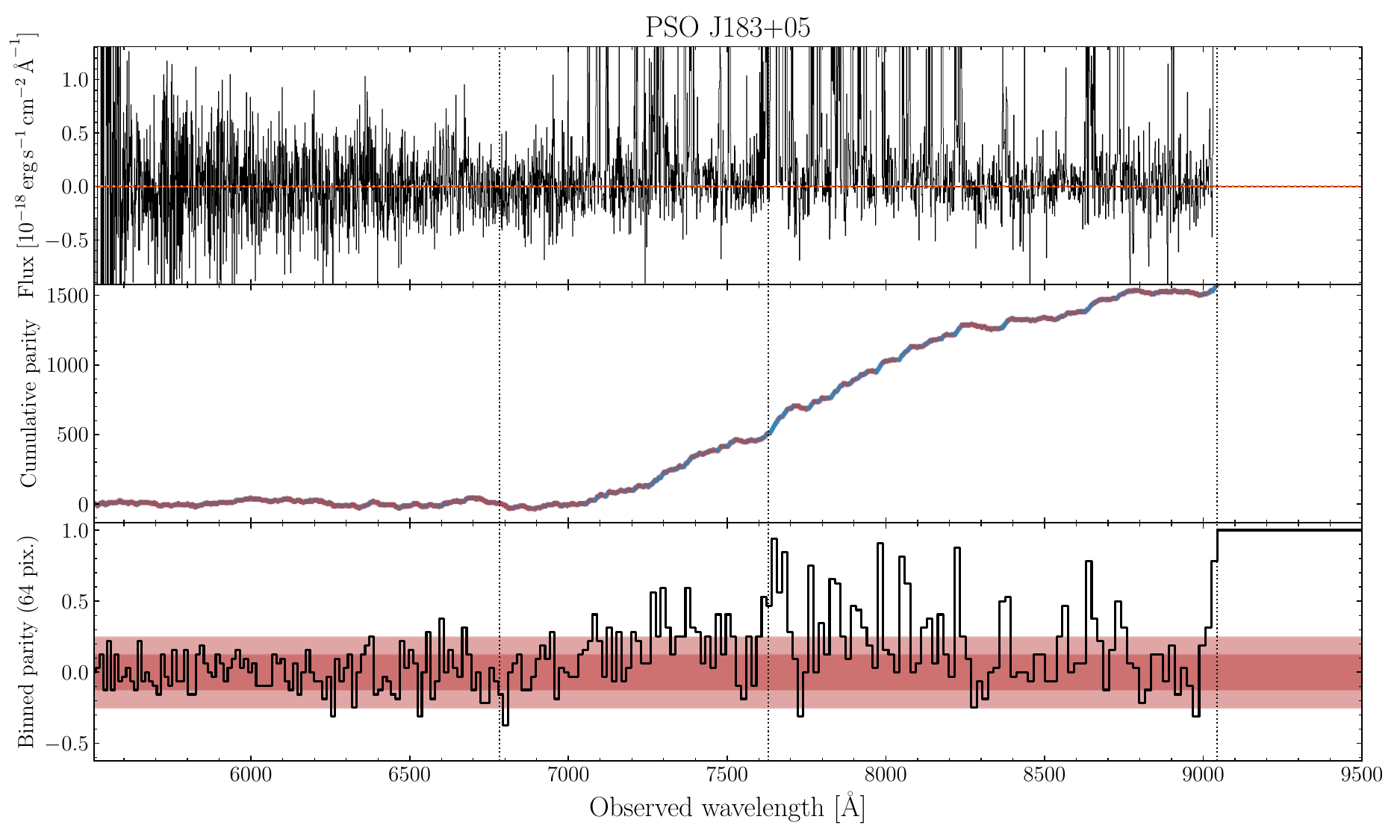}}
\end{center}
\caption{Examples of typical E-XQR-30 quasar spectra (PSO~J183$-$12 \& PSO~J183$+$05) which pass the parity tests described in the text. The top panels show the quasar spectrum zoomed in close to the zero-level (dotted orange line) after a 7-pixel median filter has been applied. The middle panels show the cumulative parity, with blue and red regions corresponding to positive and negative pixels, respectively. Regions with transmitted flux correspond to positive slope, while regions with zero flux should be roughly flat. The bottom panels show the parity averaged within 64-pixel bins, as well as dark and light shaded regions corresponding to the $1\sigma$ and $2\sigma$ scatter expected in the binned parity in regions with zero flux.}
\label{fig:good_parity}
\end{figure*}

\begin{figure*}
\begin{center}
\resizebox{16cm}{!}{\includegraphics[trim={1.0em 1em 1em 0.5em},clip]{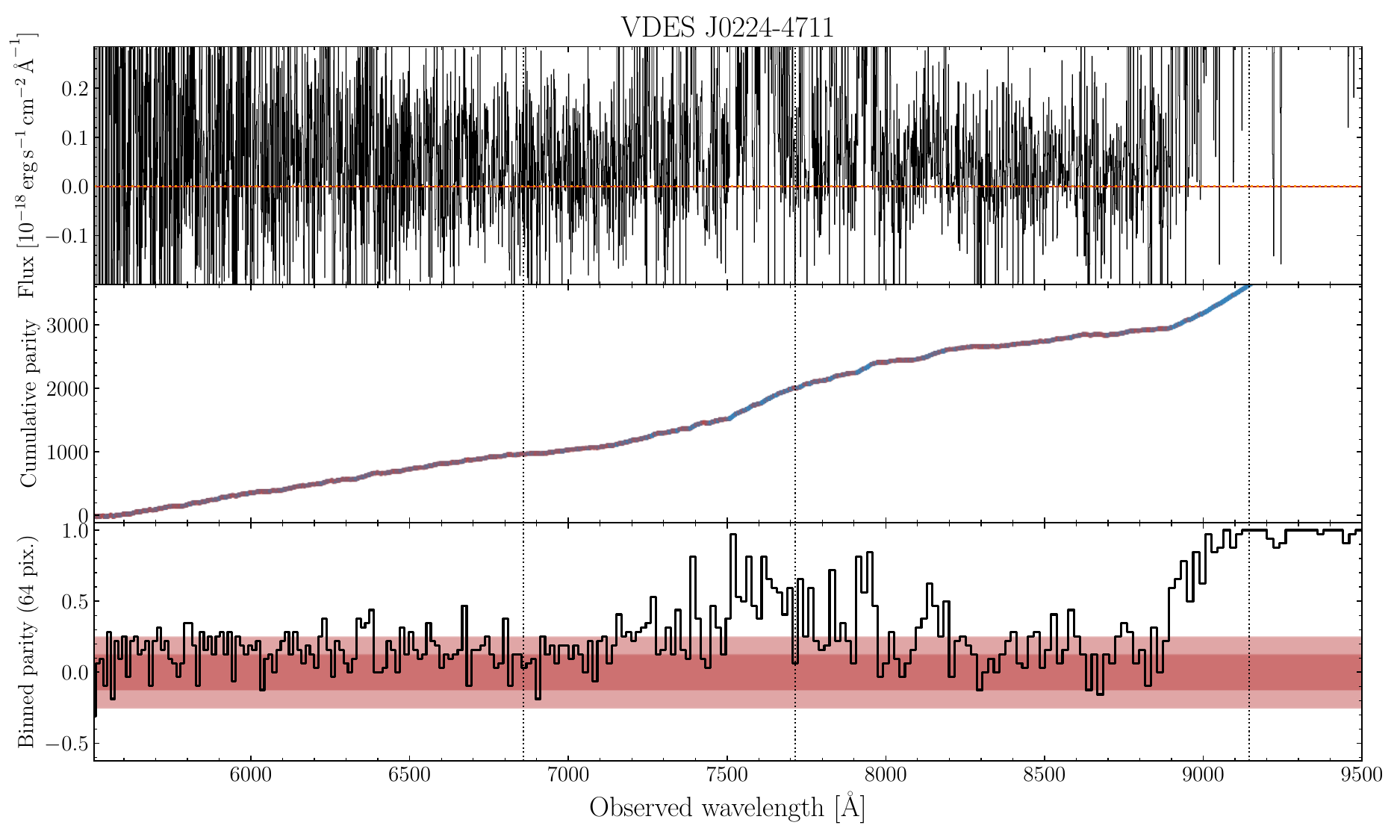}} \\
\resizebox{16.2cm}{!}{\includegraphics[trim={0.0em 1em 1em 0.5em},clip]{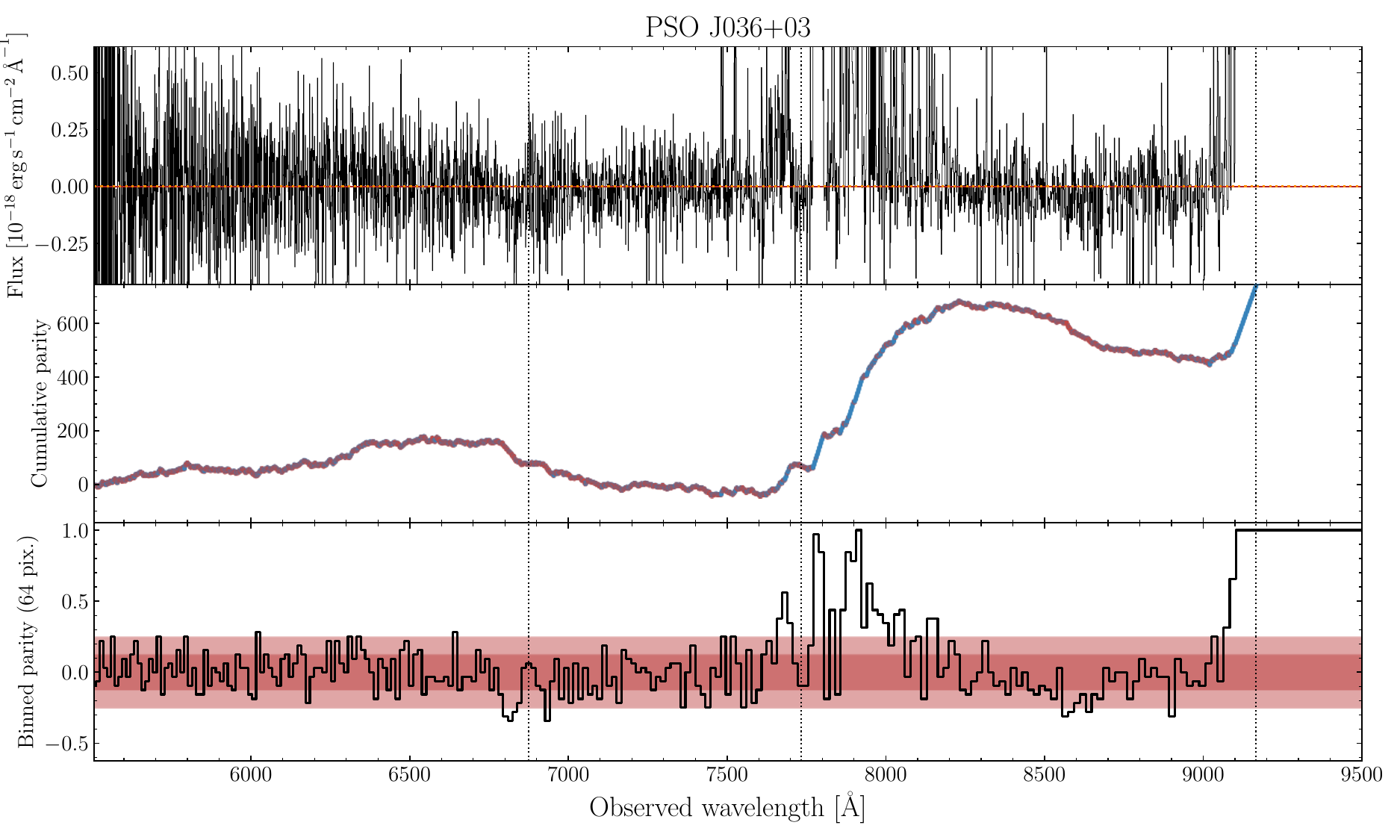}}
\end{center}
\caption{Similar to Figure~\ref{fig:good_parity} but now showing examples of quasar spectra which fail the parity tests (VDES~J0224$-$4711 \& PSO~J036$+$03).}
\label{fig:bad_parity}
\end{figure*}

As described in Section~\ref{sec:data}, we examined the positive/negative parity of the E-XQR-30 spectra to identify those with insufficiently accurate sky subtraction or other artifacts which could bias the dark fraction measurements. In Figure~\ref{fig:good_parity}, we show two examples of typical spectra which pass the parity tests. The top panels show the spectra close to the zero level after applying a 7-pixel median filter, the middle panels show the cumulative pixel parity, and the bottom panels show 64-pixel binned parity. The vertical lines show wavelengths corresponding to the Lyman limit (911.76\,\AA), Ly$\beta$ (1025.72\,\AA), and Ly$\alpha$ (1215.67\,\AA) in the rest frame of the quasar. The cumulative parity is fairly flat at the shortest wavelengths blueward of the Lyman limit, while the binned parity (in regions of zero apparent flux) shows no systematic excursions to negative or positive values.

In Figure~\ref{fig:bad_parity}, we show examples of spectra which fail the parity tests. In the top (VDES~J0224$-$4711) and bottom (PSO~J036$+$03) panels, the cumulative parity drifts upward and downward, respectively. The quasar spectrum in the top panel shows weak, but statistically significant, excess flux across the entire spectrum, leading to no negative binned pixels in all the Lyman-series forests. In the bottom panel, clear excursions to negative parity are present in dark regions across the spectrum, leading to a large excess in negative binned pixels.

 \begin{figure*}
\begin{center}
\resizebox{8.7cm}{!}{\includegraphics[trim={0.5em 1em 1em 1em},clip]{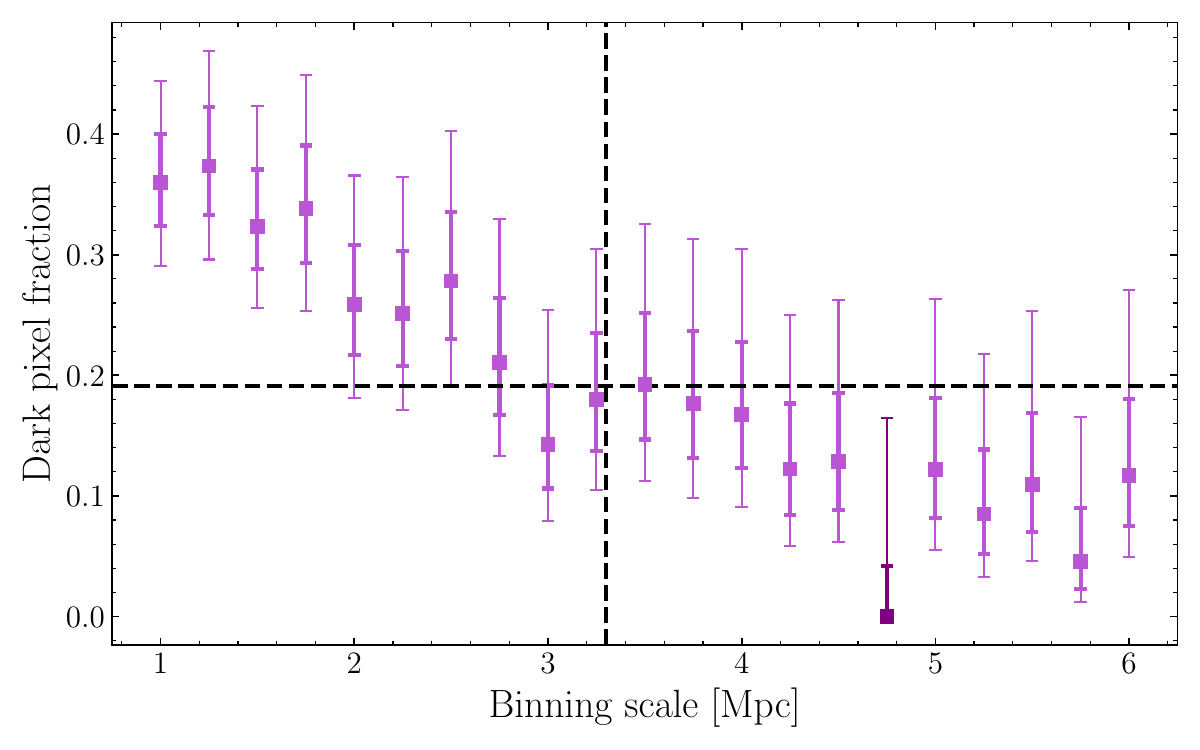}}
\resizebox{8.7cm}{!}{\includegraphics[trim={0.5em 1em 1em 1em},clip]{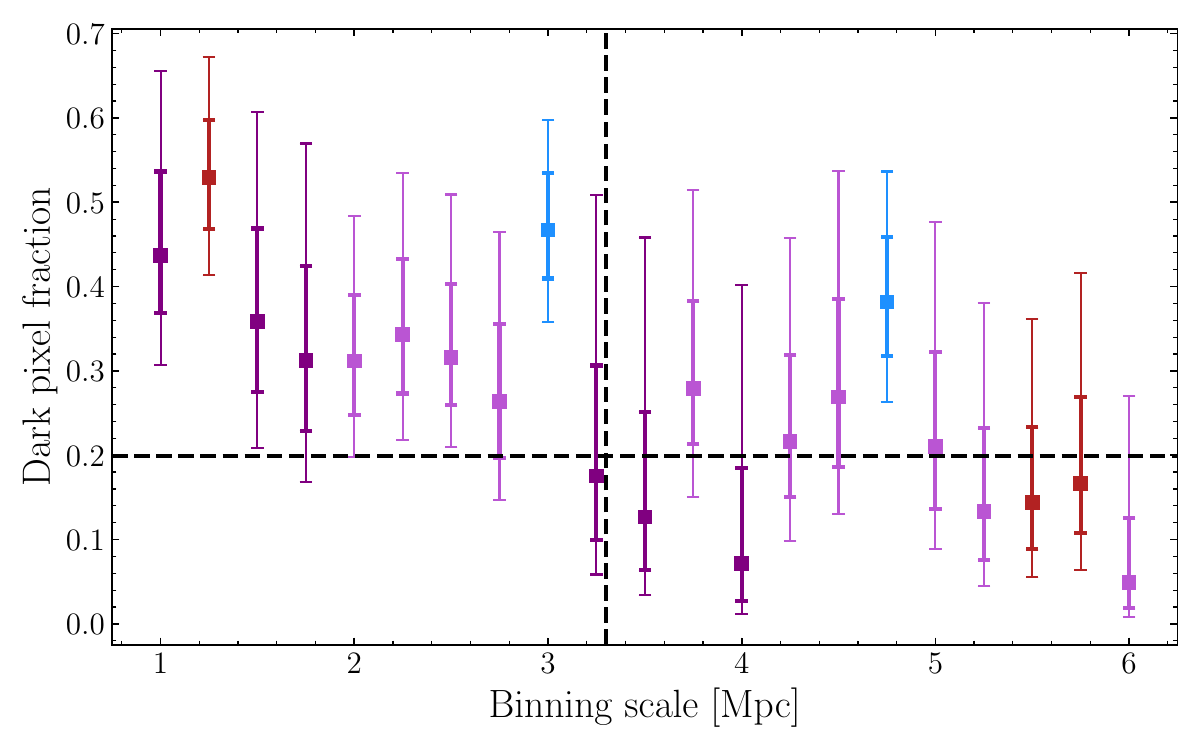}}
\end{center}
\vspace{-1.0em}
\caption{Optimal dark pixel fraction (lower upper $2\sigma$ bound) at $z\simeq5.83$ (left) and $z\simeq6.04$ (right), where the color represents the optimal forest or forest combination (similar to Figure~\ref{fig:xhi_fiducial}). The dark pixel fraction and binning scale used in this work are shown by the intersection of the black dashed lines.}
\label{fig:binscale}
 \end{figure*}

 \begin{figure*}
\begin{center}
\resizebox{11cm}{!}{\includegraphics[trim={1.0em 1em 1em 0.5em},clip]{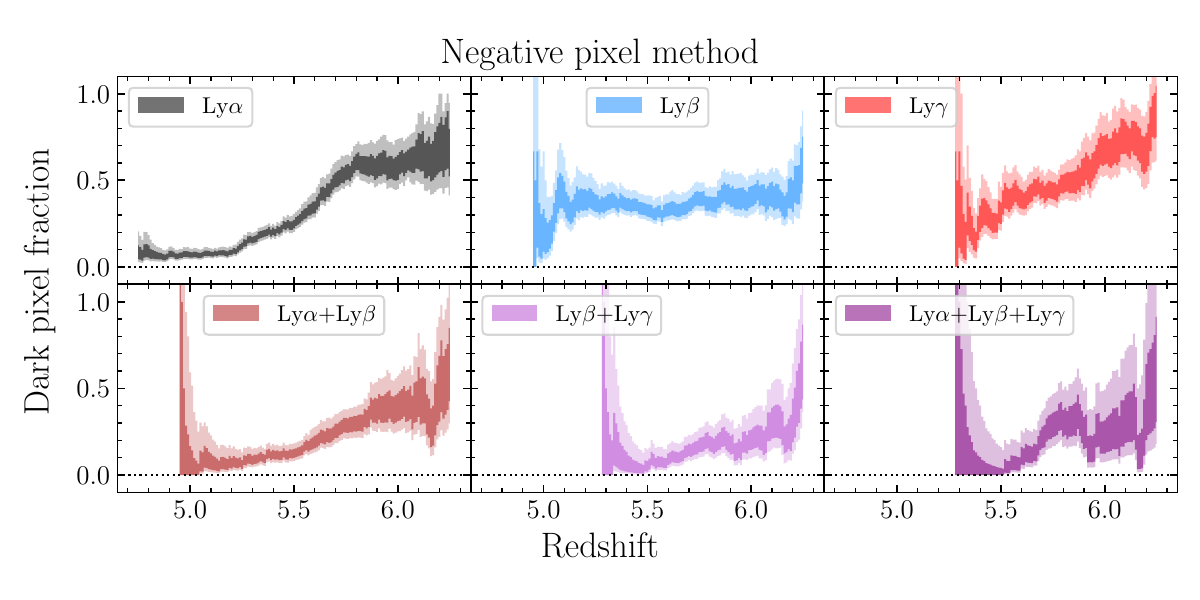}}
\end{center}
\caption{Similar to Figure~\ref{fig:dark_negpix}, but shifting the $\Delta z=0.2$ bins in increments of $\delta z=0.01$. The $1\sigma$ and $2\sigma$ constraints are shown by the dark and light shaded regions, respectively.}
\label{fig:binz}
\end{figure*}

\section{Binning tests}\label{app:bin}

In our main results, we opt for a binning scale of 3.3\,Mpc with $\Delta z=0.2$ redshift bins starting at $z=4.75$. It is worth investigating the degree to which our results depend on these choices, which introduce a degree of model dependence.

First, we examine the dependence on the binning scale. In Figure~\ref{fig:binscale}, we show how our fiducial \xhi\ constraints in the $z\simeq5.83$ (left) and $z\simeq6.04$ redshift bins (right), determined via the negative pixel method, varies with the binning scale. The dark fraction decreases more or less monotonically with increasing bin size, with some scatter due to the small number statistics involved in the negative pixel method. The resulting fiducial upper limits on the IGM neutral fraction would be $0.36+0.04$ ($0.44+0.10$) for 1\,Mpc bins and $0.12+0.06$ ($0.05+0.08$) for 6\,Mpc bins at $z\simeq5.83$ ($z\simeq6.04$). Smaller binning scales likely include more small-scale opaque regions within fully ionized regions (e.g.~\citealt{Spina24}), while larger binning scales may mix transparent ionized regions with dark neutral regions. 

Next, we look at the impact of shifting the redshift bins. In Figure~\ref{fig:binz}, we show the dark pixel fractions of every forest+combination in $\delta z=0.01$ steps, effectively a moving average over a range of $\Delta z=0.2$. While there are some fluctuations from step to step, they are largely consistent within the bootstrap uncertainties, and our specific set of redshift bins is not an outlier within the distribution.

\section{Higher order Lyman-series lines}\label{app:delta}

With the increased sensitivity of our constraints derived from the Ly$\gamma$ forest, one may wonder if pushing further down the Lyman series would be of any benefit. Only the next Lyman series transition, Ly$\delta$ ($\lambda_\delta = 949.74$\,\AA), has any coverage redward of the Lyman limit after our fiducial proximity zone exclusion, but with only $\sim1/3$ of the path length of Ly$\gamma$ (see Figure~\ref{fig:npix}) there are relatively few pixels with which to constrain the dark pixel fraction. Here we examine whether there is any additional constraining power to be obtained from the Ly$\delta$ forest.

\begin{table*}\centering
\begin{tabular}{c c c c c c c c c c c}
Line & $z$ range & $\bar{z}$ & $N_{\rm LOS}$ & $N_{\rm pix}$ & $N_{\rm un-det}$ & $f_{\rm dark}^{\rm thresh}+1\sigma/+2\sigma$ & $N_{\rm neg}$ & $f_{\rm dark}^{\rm neg}+1\sigma/+2\sigma$ \\
\hline \hline
Ly$\delta$ & 5.55--5.75 & 5.639 & 15 & 67 & 34 & $0.519+0.100(+0.197)$ & 14 & $0.418+0.108(+0.229)$ \\
& 5.75--5.95 & 5.828 & 19 & 127 & 87 & $0.701+0.061(+0.124)$ & 30 & $0.472+0.079(+0.165)$ \\
& 5.95--6.15 & 6.056 & 7 & 52 & 38 & $0.747+0.096(+0.182)$ & 15 & $0.577+0.133(+0.274)$ \\
& 6.15--6.35 & 6.226 & 3 & 23 & 23 & $1.023+0.000(+0.000)$ & 14 & $1.217+0.211(+0.383)$ \\
\hline Ly$\gamma$+Ly$\delta$ & 5.55--5.75 & 5.639 & 15 & 65 & 19 & $0.306+0.082(+0.179)$ & 5 & $0.308+0.159(+0.373)$ \\
& 5.75--5.95 & 5.828 & 19 & 125 & 55 & $0.460+0.059(+0.118)$ & 6 & $0.192+0.092(+0.217)$ \\
& 5.95--6.15 & 6.056 & 7  & 50  & 30 & $0.628+0.120(+0.244)$ & 4 & $0.320+0.191(+0.445)$ \\
& 6.15--6.35 & 6.226 & 3  & 23  & 17 & $0.773+0.074(+0.273)$ & 6 & $1.043+0.401(+0.832)$ \\
\hline Ly$\beta$+Ly$\gamma$+Ly$\delta$ & 5.55--5.75 & 5.639 & 15 & 64 & 10 & $0.167+0.080(+0.178)$ & 1 & $0.125+0.196(+0.559)$ \\
& 5.75--5.95 & 5.828 & 19 & 125 & 31 & $\mathbf{0.265+0.053(+0.109)}$ & 2 & $0.128+0.125(+0.334)$ \\
& 5.95--6.15 & 6.056 & 7 & 49 & 20 & $0.437+0.175(+0.339)$ & 1 & $0.163+0.254(+0.714)$ \\
& 6.15--6.35 & 6.226 & 3 & 23 & 9  & $0.419+0.235(+0.651)$ & 3 & $1.043+0.687(+1.564)$ \\
\hline Ly$\alpha$+Ly$\beta$+Ly$\gamma$+Ly$\delta$ & 5.55--5.75 & 5.639 & 15 & 64 & 9 & $\mathbf{0.154+0.073(+0.162)}$ & 0 & $0.000+0.246(+0.941)$ \\
& 5.75--5.95 & 5.828 & 19 & 124 & 31 & $0.274+0.052(+0.109)$ & 1 & $0.129+0.206(+0.599)$ \\
& 5.95--6.15 & 6.056 & 7  & 46  & 16 & $0.381+0.167(+0.349)$ & 1 & $0.348+0.540(+1.512)$ \\
& 6.15--6.35 & 6.226 & 3  & 23  & 8  & $0.381+0.276(+0.714)$ & 0 & $0.000+0.667(+2.370)$ \\
\hline
\end{tabular}
\caption{Dark pixel statistics from the Ly$\delta$ forest of the E-XQR-30 sample, similar to Table~\ref{tab:results}. Uncertainty ranges for $f_{\rm dark}$ can go above unity due to the re-scaling factors applied to convert the corresponding fractions of un-detected or negative pixels.}
\label{tab:results_delta}
\end{table*}

\begin{figure*}
\begin{center}
\resizebox{14cm}{!}{\includegraphics[trim={0.5em 1em 1em 1em},clip]{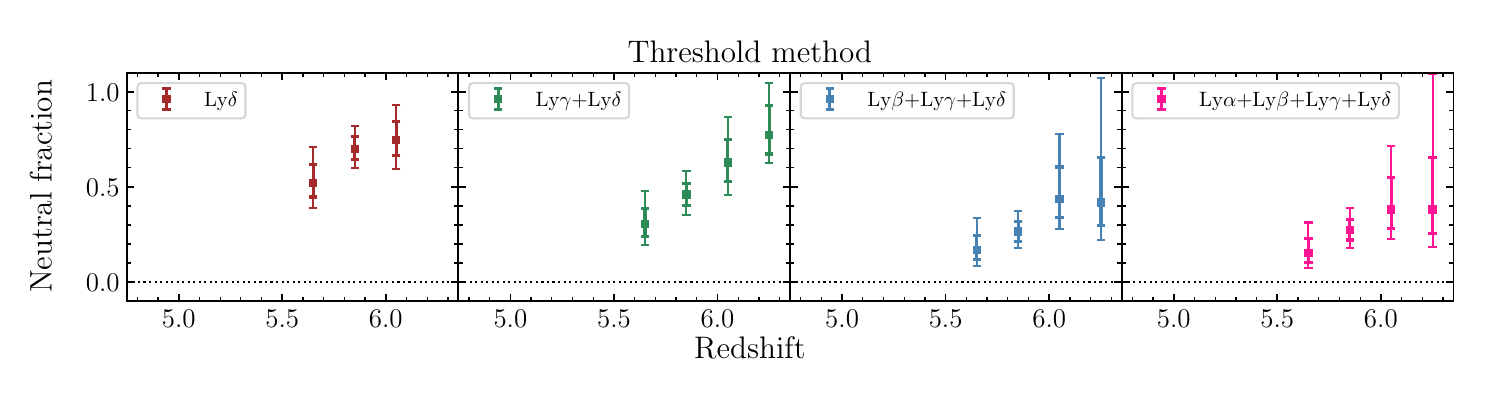}}\\
\resizebox{14cm}{!}{\includegraphics[trim={0.5em 1em 1em 1em},clip]{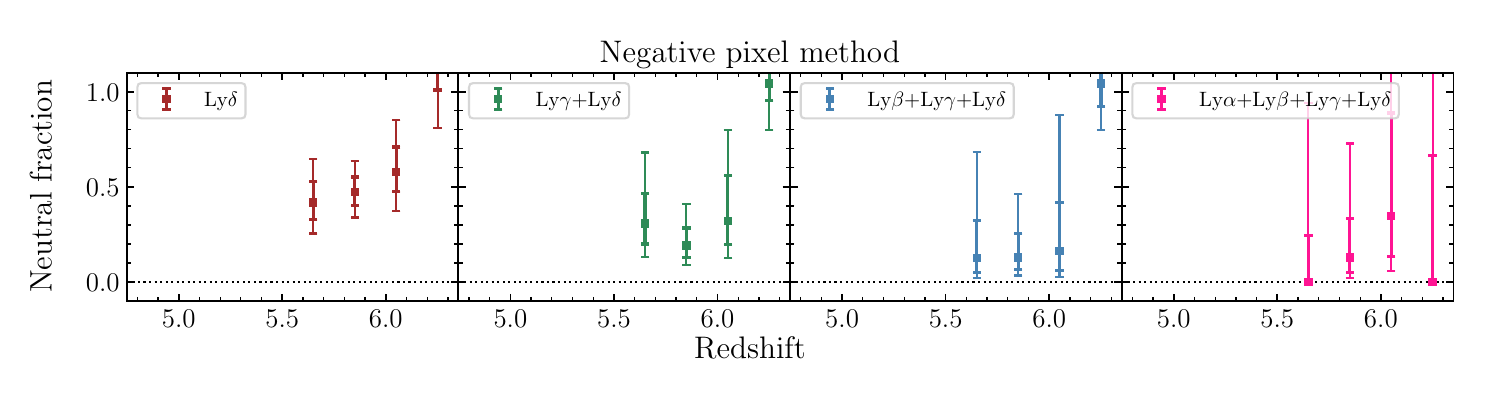}}
\end{center}
\caption{Similar to Figure~\ref{fig:dark_thresh} but for the Ly$\delta$ forest and its combinations.}
\label{fig:delta}
 \end{figure*}

\begin{figure*}
\begin{center}
\resizebox{8.7cm}{!}{\includegraphics[trim={1.0em 1em 1em 0.5em},clip]{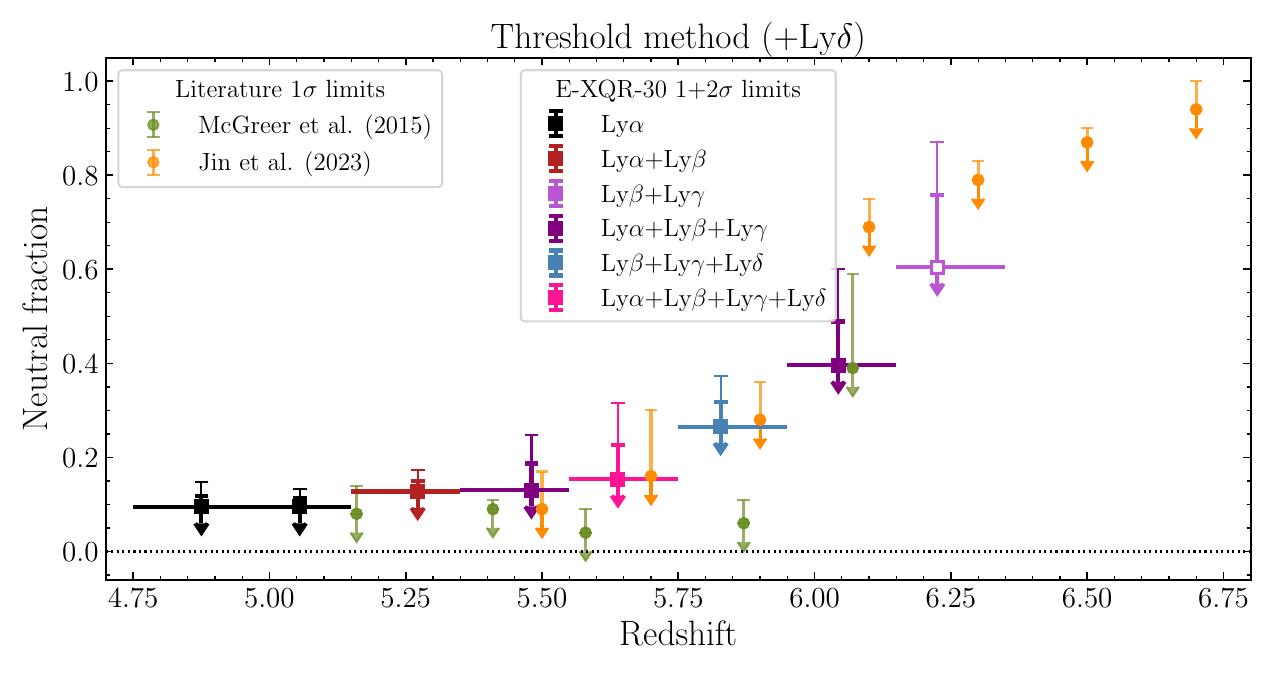}}
\end{center}
\caption{Similar to Figure~\ref{fig:xhi_fiducial} but now including the more sensitive upper limits at $z=5.55$--$5.75$ from the Ly$\alpha$$+$Ly$\beta$$+$Ly$\gamma$$+$Ly$\delta$ forest and $z=5.75$--$5.95$ from the Ly$\beta$$+$Ly$\gamma$$+$Ly$\delta$ forest.}
\label{fig:xhi_delta}
\end{figure*}

In Figure~\ref{fig:delta}, we show the dark pixel fractions obtained from the Ly$\delta$, Ly$\gamma$+Ly$\delta$, Ly$\beta$+Ly$\gamma$+Ly$\delta$, and Ly$\alpha$+Ly$\beta$+Ly$\gamma$+Ly$\delta$ forests\footnote{For simplicity, as in our fiducial results, we neglect constraints that skip an intermediate Lyman-series line.} (see also Table~\ref{tab:results_delta}). While the dark fractions from the Ly$\delta$ forest combinations are among the smallest at $z\sim6$, due to the small number of pixels they generally suffer from large statistical uncertainty, particularly when used with the negative pixel method. That said, in two cases, namely the $z\simeq5.65$ and (just barely) $z\simeq5.85$ bins in the threshold method, the combination of Ly$\alpha$+Ly$\beta$+Ly$\gamma$+Ly$\delta$ and Ly$\beta$+Ly$\gamma$+Ly$\delta$ provide tighter $2\sigma$ upper limits, respectively, and thus could take the place of the fiducial upper limits at those redshifts in the left panel of Figure~\ref{fig:xhi_fiducial}. We err on the side of caution and quote the Ly$\alpha$+Ly$\beta$+Ly$\gamma$ constraints as our fiducial threshold method limits in these redshift bins, but show the corresponding constraints in Figure~\ref{fig:xhi_delta} and the bold entries in Table~\ref{tab:results_delta}.

Our exploration of Ly$\delta$ suggests that the higher order Lyman-series can be more sensitive, but will require many more quasar spectra to be statistically sound. Fainter quasars, with correspondingly smaller proximity zones, also have the potential to open up significant additional path length. But while the proximity zone size should scale with $L_{\rm qso}^{-1/2}$, the required exposure time to reach comparable sensitivity (in the background-limited regime) scales as $L_{\rm qso}^{-2}$, making such an experiment rather costly.


\bsp	
\label{lastpage}
\end{document}